\documentclass[twocolumn,twocolappendix]{aastex631}

\usepackage{graphicx}	% Including figure files
\usepackage{amsmath}	% Advanced maths commands
\usepackage{amssymb}	% Extra maths symbols
\usepackage{xcolor}
\usepackage{ctable}
\usepackage{booktabs}
\usepackage{threeparttable}
\usepackage{comment}
\usepackage{newtxtext,newtxmath}

\usepackage[T1]{fontenc}
\usepackage{hyperref}

\DeclareRobustCommand{\VAN}[3]{#2}
\let\VANthebibliography\thebibliography
\def\thebibliography{\DeclareRobustCommand{\VAN}[3]{##3}\VANthebibliography}
\graphicspath{ {./figures/} }

\begin{document}

\title{Baryon Pasting the Uchuu Lightcone Simulation}

\correspondingauthor{}
\author[0000-0001-8914-8885]{Erwin T. Lau}
\email{erwin.lau@cfa.harvard.edu}
\affiliation{Center for Astrophysics $\vert$ Harvard \& Smithsonian, Cambridge, MA, 02138, USA}

\author[0000-0002-6766-5942]{Daisuke Nagai}
\affiliation{Department of Physics, Yale University, New Haven, CT 06520, USA}

\author[0000-0003-0777-4618]{Arya~Farahi}
\affiliation{Department of Statistics and Data Sciences, University of Texas at Austin, Austin, TX 78712, USA}
\affiliation{The NSF-Simons AI Institute for Cosmic Origins, University of Texas at Austin, Austin, TX 78712, USA}

\author[0000-0002-5316-9171]{Tomoaki Ishiyama}
\affiliation{Digital Transformation Enhancement Council, Chiba University, Chiba 263-8522, Japan}

\author[0000-0001-7964-9766]{Hironao Miyatake}
\affiliation{Kobayashi-Maskawa Institute for the Origin of Particles and the Universe (KMI), Nagoya University, Nagoya, 464-8602, Japan}
\affiliation{Institute for Advanced Research, Nagoya University, Nagoya 464-8601, Japan}
\affiliation{Kavli Institute for the Physics and Mathematics of the Universe (WPI), The University of Tokyo, Kashiwa, Chiba 277-8583, Japan}
\affiliation{The University of Tokyo Institutes for Advanced Study (UTIAS), The University of Tokyo, Chiba 277-8583, Japan} 

\author[0000-0002-7934-2569]{Ken Osato}
\affiliation{Center for Frontier Science, Chiba University, Chiba 263-8522, Japan}
\affiliation{Department of Physics, Graduate School of Science, Chiba University, Chiba 263-8522, Japan}
\affiliation{Kavli Institute for the Physics and Mathematics of the Universe (WPI), The University of Tokyo, Kashiwa, Chiba 277-8583, Japan}

\author[0000-0002-1706-5797]{Masato Shirasaki}
\affiliation{National Astronomical Observatory of Japan (NAOJ), National Institutes of Natural Science, Mitaka, Tokyo 181-8588, Japan}
\affiliation{The Institute of Statistical Mathematics,
Tachikawa, Tokyo 190-8562, Japan}

\begin{abstract}
We present the Baryon Pasted (BP) X-ray and thermal Sunyaev-Zel'dovich (tSZ) maps derived from the half-sky Uchuu Lightcone simulation. These BP-Uchuu maps are constructed using more than $75$ million dark matter halos with masses $M_{500c} \geq 10^{13} M_\odot$ within the redshift range $0 \leq z \leq 2$. A distinctive feature of our BP-Uchuu Lightcone maps is their capability to assess the influence of both extrinsic and intrinsic scatter caused by triaxial gaseous halos and internal gas characteristics, respectively, at the map level. We show that triaxial gas drives substantial scatter in X-ray luminosities of clusters and groups, accounting for nearly half of the total scatter in core-excised measurements. Additionally, scatter in the thermal pressure and gas density profiles of halos enhances the X-ray and SZ power spectra, leading to biases in cosmological parameter estimates. These findings are statistically robust due to the extensive sky coverage and large halo sample in the BP-Uchuu maps. The BP-Uchuu maps are publicly available online via \href{https://app.globus.org/file-manager?origin_id=cf8dadb7-b6e9-4e2c-abc1-0813877efc13}{Globus}. 
\end{abstract}

\keywords{Large-scale structure of Universe, Galaxy clusters, Intracluster Medium, N-body simulations}

\section{Introduction}

Ongoing and upcoming large-scale multiwavelength sky surveys in X-ray, microwave, and optical of galaxy clusters are promising for improving our understanding of cosmology. This is going to be achieved with significant reduction of statistical uncertainties owing to the large cosmological volumes that these surveys will probe. In particular, the order of magnitude increase in the number of galaxy clusters and groups compared to cluster surveys in the previous decade will dramatically improve the cosmological constraints of cluster abundance measurements \citep[see][for a review]{allen_etal11}. The extensive sky coverage of these surveys also enables cross-correlations of clusters and groups as powerful cosmological and astrophysical probes \citep[e.g.,][]{shirasaki_etal20}. 
The success of these surveys depends on our ability to accurately model and mitigate systematic uncertainties, which can profoundly influence the distribution and evolution of matter on small to intermediate scales.

Accurate estimation of the mass-observable scaling relations is fundamental to derive reliable constraints on key cosmological parameters in cluster abundance cosmology \citep{pratt_etal19}.  These scaling relations connect the halo mass to observable properties, such as X-ray luminosity or the thermal Sunyaev--Zeldovich (tSZ) signal at microwave wavelengths, enabling indirect mass estimation in cosmological surveys \citep[e.g.,][]{vikhlinin_etal09, mantz2010observed, benson_etal13, bocquet2019cluster}. However, these relations often fail to capture the full complexity of physical systems because they do not account for the scatter around the mean relations \citep[e.g.,][]{mantz2016weighing, sereno2020xxl}. While scatter provides direct information on the structure and evolution of these systems \citep{farahi2019NatureCorrelation}, failing to account for scatter biases in the interpretation of observational data \citep{farahi2019mass,costanzi2019methods, zhang2024bias}, thereby affecting the precision estimates of crucial cosmological parameters such as the matter density ($\Omega_m$) and the amplitude of matter fluctuations ($\sigma_8$).

Scatter in scaling relations can be categorized into two primary sources: intrinsic and extrinsic. Intrinsic scatter arises from internal physical processes within the halo, such as baryonic physics such as feedback from supernovae (SNe) and active galactic nuclei (AGN), radiative gas cooling, and star formation, all of which influence the thermodynamics of the halo gas \citep[e.g.,][]{farahi_etal18, Truong2018scatter_simulations, pop_etal22, yang_etal22}. Extrinsic scatter, on the other hand, originates from external factors such as the non-spherical (triaxial) nature of halos and projection effects. Similarly to dark matter (DM) distributions, the gas distributions in clusters and groups are typically triaxial \citep{lau_etal11, mulroy2019locuss, kim_etal24}, and the observed signals can be affected by the orientation of the halo's major axis relative to the line of sight \citep[e.g.,][]{battaglia_etal12, limousin2013three}. Understanding and quantifying both intrinsic and extrinsic scatter is essential to improve the accuracy of mass estimates and minimizing biases in cosmological inferences.

To date, extrinsic scatter in cosmological inferences remains incompletely quantified. Current simulations often lack the resolution or comprehensive physical models necessary to accurately capture the interplay between various baryonic processes that contribute to correlated and extrinsic scatter. They are also expensive to run, making them less suitable for exploring the wide parameter spaces required for upcoming large-scale multiwavelength surveys, where numerous combinations of cosmological and astrophysical parameters must be tested \citep{mead2021hmcode, camels, flamingo}. However, observational constraints on these types of scatter are limited as it is difficult to separate extrinsic scatter from intrinsic scatter with observations. This makes it challenging for observations to validate and refine simulation predictions of intrinsic scatter. The inability to efficiently explore these parameter spaces hampers our capacity to fully understand and mitigate the systematic uncertainties introduced by baryonic physics in cosmological analyses.

A complementary and more feasible approach involves the use of empirical and analytic methods to model baryonic properties within DM halos \citep[see][for a review]{wechsler_tinker18}. This strategy entails forward-modeling baryonic properties by overlaying them onto DM halos from relatively inexpensive large-scale DM-only cosmological simulations. These baryon-DM models are calibrated using either detailed cosmological simulations or high-quality observational datasets. By varying the underlying baryon-DM models and cosmological parameters within DM-only simulations, one can efficiently study the dependence of baryonic effects on astrophysical processes and cosmology. This empirical and analytic framework not only facilitates the exploration of a broader parameter space but also supports the statistical inference necessary to maximize the scientific return of upcoming surveys. Moreover, it allows for rapid iterations and refinements based on new observational data, ensuring that the models remain accurate and relevant in the rapidly evolving landscape of cosmological research.

Although significant efforts have been dedicated to painting galaxy properties onto DM halos \citep[e.g.,][]{seljak2000analytic,moster2018emerge, behroozi2019universemachine,hadzhiyska2020limitations}, fewer works focus on painting gas properties. Existing gas painting techniques in the literature \citep{valotti2018cosmological, clerc2018synthetic, zandanel_etal18, websky, comparat_etal20, agora, williams_etal23, bayer_etal24} rely on phenomenological or empirical gas models calibrated with cosmological hydrodynamical simulations or observations. One key disadvantage of this approach is that empirical parameters often lack direct physical interpretation, making them less useful for interpreting observations or for understanding the underlying physics. 

Alternatively, advances in machine learning techniques have been applied to paint galaxies or gas on DM halos \citep[e.g.,][]{agarwal2018painting, chadayammuri_etal23, nguyen2024dreams}. However, these approaches are often ``black boxes'', making them difficult to interpret physically. Hybrid approaches \citep[e.g.,][]{picasso}, calibrate parameters of gas analytical models using machine learning, offering more physically interpretable models. Nevertheless, they still rely on large training samples of input simulations that are expensive to produce and are subject to the same subgrid physics uncertainties inherent in the simulations they are trained on. 

To address the challenges of modeling halo gas for large-scale multiwavelength galaxy surveys, we have developed Baryon Pasting (BP), a code specifically designed to paint gas onto DM halos efficiently and fast, with physically motivated models. BP provides physically interpretable modeling of gas in DM halos, which is essential for astrophysical and cosmological inferences from observations. It enables rapid exploration of how feedback physics \citep{ostriker_etal05,bode_etal09,trac_etal11} and non-thermal pressure support affect the tSZ power spectrum \citep{shaw_etal10}, how cool-dense cores affect optical depth measurements of clusters and groups in kinetic SZ observations \citep{flender_etal17}, and how to break the degeneracies between astrophysical and cosmological parameters in cross-correlations of X-ray, tSZ, and lensing measurements of galaxy clusters and groups \citep{shirasaki_etal20}. Painting gas on DM particles in the simulation allows us to model gas observables in cosmic web filaments \citep{bp_algo}. Unlike purely empirical or machine learning-based methods, BP maintains physical interpretability while remaining computationally efficient, making it well-suited for the extensive parameter space exploration required by upcoming surveys. 

In this paper, we present the updated BP gas model and apply it to create half-sky multi-wavelength maps using the DM-only Uchuu Lightcone simulations \citep{uchuu}. Specifically, these BP-Uchuu mock maps enable quantification of the impact of halo triaxiality on the X-ray luminosity ($L_X$)–mass scaling relation, focusing specifically on extrinsic scatter. We classify scatter due to triaxiality and projection as extrinsic, differentiating it from intrinsic scatter caused by internal halo processes. Our results show that halo orientation plays a significant role in modulating observed X-ray luminosity, with halos aligned along the line of sight exhibiting enhanced luminosities. Specifically, halos with their elongated axis more aligned with the line-of-sight direction tend to show systematically higher $L_X$ values for a given mass. The comparison between triaxial and spherical halos reveals that this bias is not merely a projection effect, but rather a result of halo shape and orientation, contributing an additional scatter to the $L_X$ measurements. This extrinsic scatter is non-negligible at the $8\%-9\%$ level, at around 1/3 and 1/2 of the total scatter, for non-core-excised and core-excised $L_X$, respectively. Our results further highlight the importance of accounting for halo triaxiality in precision cosmological analyses. In addition, we also show that the scatter in the halo gas profiles leads to $\sim 10\%-40\%$ biases in the X-ray and tSZ angular power spectra of clusters and groups \citep{hurier_etal15, planck_tsz, lau_etal23, lau_etal24}. Interpretation of the power spectrum that uses the halo model usually does not include intrinsic scatter in halo gas profiles, leading to underestimates of the actual amplitude of the power spectrum and thus biases the derived constraints on cosmological parameters such as $\sigma_8$ and $\Omega_m$. 
 
We give an overview of the BP gas model in Section~\ref{sec:model} and how the BP maps are generated in Section~\ref{sec:bp_maps}. We present results with the Uchuu BP maps in Section~\ref{sec:science}. Finally, we give our conclusions in Section~\ref{sec:summary}. 

\section{Gas Model}\label{sec:model}

\begin{table*}
\begin{center}
\begin{tabular}{|c|l|l|l|}
\hline
Parameter &  Physical meaning & Equation & Default Value \\
\hline \hline
$\epsilon_f$ & feedback efficiency from SNe and AGN & Eq.~(\ref{eq:energy})& $3.97 \times 10^{-6}$ \\
$\epsilon_{\rm DM}$ & DM energy transfer to gas & Eq.~(\ref{eq:energy}) & $0.0$  \\
$f_\star$ & amplitude of the stellar mass fraction  & Eq.~(\ref{eq:fstar}) & $0.026$ \\
$S_\star$ & mass slope of stellar mass fraction & Eq.~(\ref{eq:fstar}) & $0.12$ \\
$x_{\mathrm{break}} $ & cluster core radius in $R_{500}$ & Eq.~(\ref{eq:polytropic_index}) & $0.195$  \\
$\Gamma $ & polytropic index outside the cluster core & Eq.~(\ref{eq:polytropic_index}) & $1.2$ \\
$\Gamma_0 $ & polytropic index within cluster core  & Eq.~(\ref{eq:polytropic_index}) & $0.1024$ \\
$\beta_g $ & redshift evolution of polytropic index within cluster core & Eq.~(\ref{eq:polytropic_index}) & $1.72$ \\
$A_{\rm nt}$ & amplitude of non-thermal pressure fraction profile & Eq.~(\ref{eq:nelson}) & 0.452 \\
$B_{\rm nt}$ & scale of the radial dependence of the non-thermal pressure fraction profile in $R_{200m}$ & Eq.~(\ref{eq:nelson}) & 0.841 \\
$\gamma_{\rm nt}$ & logarithmic slope of non-thermal pressure fraction profile & Eq.~(\ref{eq:nelson}) & 1.628 \\
$R_{\rm bound}$ & Boundary of the gas in halo in DM splashback radius $R_{\rm sp}$ & -- & 1.89 \\
\hline
\end{tabular}
\end{center}
\caption{\label{tab:param} Default values of the Gas model parameters adopted by our BP gas model. They are taken from the best fits of the gas density profiles from {\em Chandra}-SPT cluster samples from \citet{flender_etal17}, except the parameters for the non-thermal pressure fraction profile and the clumping profile, which are taken from the fit to the {\em Omega500} cosmological simulation \citep{nelson_etal14}. }
\end{table*}

\begin{figure}
\includegraphics[width=1.0\columnwidth]{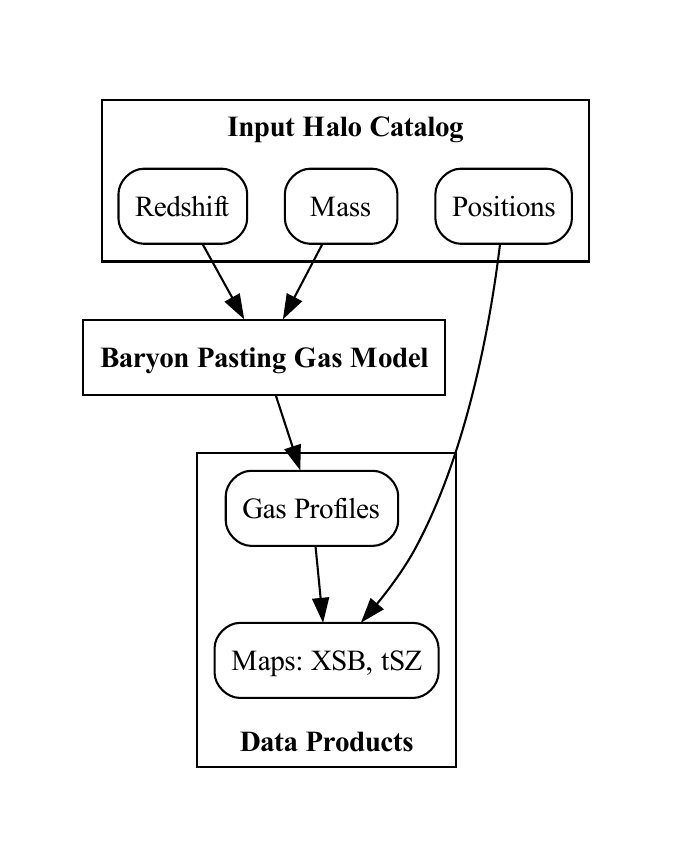} 
\vspace{-1.2cm}
    \caption{A Schematic diagram showing the general structure of BP map-making. We start with an input halo catalog with information on the redshift, mass (specifically $M_{\rm vir}$), and positions (RA and Dec) of the halo. Given the halo mass and redshift, we generate the gas profiles (X-ray emissivity, Compton-$y$) using the BP gas model. The profiles are then painted onto the X-ray Surface Brightness (XSB) and thermal Sunyaev-Zeldovich (tSZ) maps using the position information. 
    }
    \label{fig:bp_structure}
\end{figure}

Figure~\ref{fig:bp_structure} illustrates the general structure of how BP maps are generated from halo catalogs. We take 4 main halo information: redshift, mass (specifically virial mass), right ascension (RA) and declination (Dec) from the halo catalog. Optionally, we also take the halo shape information: axis ratios and the angle between the major axis and the line-of-sight for generating maps with triaxial halos. The redshift and virial mass then serve as inputs to the BP gas model to generate
gas profiles in density, pressure, temperature, Compton-$y$ profile, and X-ray emissivity profile for each halo. The gas profiles also serve as one of the primary data products from which other observables, such as scaling relations and power spectra, are derived.  To generate the map, for each halo we paint the profile (Compton-$y$ profile or the X-ray emissivity profile) at the corresponding RA and Dec of the halo on the map. 

\subsection{Halo Gas Profiles}

Here we briefly review the salient features of the core BP gas model. The BP gas model builds on a simple feedback model by \citet{ostriker_etal05}, which was then extended to include additional physical effects, such as non-thermal pressure \citep{shaw_etal10} and cool-cores \citep{flender_etal17}. We refer the reader to these papers for details. The basic assumption of the BP model is that the DM halo follows the profile Navarro--Frenk--White (NFW) profile \citep{nfw96},  
\begin{equation}
\rho_{\mathrm{NFW}}(r) =  \frac{\rho_s}{(r/r_s) (1+r/r_s)^{2}}\label{eq:nfw}
\end{equation}
where $r_s$ is the NFW scale radius and $\rho_s$ is the normalization,
which is completely specified by the virial mass of the halo $M_{\rm vir}$ and the halo concentration parameter $c_{\rm vir}$, which is defined as the ratio of the virial radius $R_{\mathrm{vir}}$ to the scale radius $r_s$ where the logarithmic slope of the DM density is $-2$. The virial radius is the radius of the sphere enclosing the virial mass, $M_{\mathrm{vir}} = \frac{4}{3} \pi R_{\mathrm{vir}}^3\Delta_{\mathrm{vir}} \rho_c(z)$ with $\Delta_{\mathrm{vir}} = 18 \pi^2 + 82 (\Omega_M(z)-1) -39 (\Omega_M(z)-1)^2$ \citep{bryan_norman98}, and $\rho_c(z)$ is the critical density at redshift $z$. 

The BP gas model assumes the {\em total} gas pressure (thermal + non-thermal) $P_{\rm{tot}}$ is in hydrostatic equilibrium (HSE) with the gravitational potential of the halo, and that the relationship between $P_{\rm{tot}}$ and the gas density $\rho_{g}$ is related through the polytropic relation, 
\begin{eqnarray}
P_{\rm{tot}}(r) &=& P_0 \theta(r)^{n+1}, \label{eq:theta_p}\\
\rho_{g}(r) &=& \rho_0 \theta(r)^{n},\label{eq:theta_rho}\\
\theta(r) &=& 1 + \frac{\Gamma-1}{\Gamma}\frac{\rho_0}{P_0}(\Phi(r=0) - \Phi(r)),\label{eq:theta} 
\end{eqnarray}
where $\theta$ is a dimensionless function that represents the gas temperature, $\Phi$ is the gravitational potential of the halo given by the NFW profile, 
\begin{equation}\label{eq:NFW_potential}
    \Phi_{\rm NFW} = -\frac{4\pi G \rho_s r_s^3 }{r}\ln\left(1+\frac{r}{r_s}\right),
\end{equation}
and ${\Gamma=1+1/n} = 1.2 $ and $n=5$ are the polytropic exponent and the polytropic index respectively, whose values are set to match those in cosmological hydrodynamical simulations \citep{komatsu_seljak01, ostriker_etal05,shaw_etal10,battaglia_etal12}. The values are also consistent with recent observations of the polytropic index \citep{ghirardini_etal19}. 

Following \citet{flender_etal17}, we model the gas in the cores of halos differently with different polytropic equations of state than the rest of the halo. Due to strong cooling and feedback in the core, the gas in the core is denser and colder, thus we adopt a smaller polytropic exponent than the outer region:
\begin{equation}\label{eq:polytropic_index}
\Gamma = 
\begin{cases}
    1.2 & r_{\rm break}/R_{500c} \geq x_{\rm break}, \\
    \Gamma_0 (1+z)^{\beta_g}  & r_{\rm break}/R_{500c} < x_{\rm break}, 
\end{cases}
\end{equation} 
with $ x_{\rm break}=r_{\rm break}/R_{500c} $ representing the radial extent of the halo core in units of $R_{500c}$ of the halo. Following \cite{flender_etal17}, we set $\Gamma_0 = 0.1024$, $\beta_g = 1.72$, and $x_{\rm break} = 0.195$. 

The normalization constants $P_0$ and $\rho_0$ are determined numerically by solving the energy and momentum conservation equations of the gas. In particular, the energy of the gas is given by
\begin{equation}\label{eq:energy}
E_{g,f}  =  E_{g,i} + \epsilon_{\rm DM} |E_{\rm DM }| + \epsilon_f M_\star c^2 + \Delta E_p.
\end{equation}
where $E_{g,f}$ and $E_{g,i}$ are the final and initial total energies (kinetic plus thermal plus potential) of the ICM. $\Delta E_p$ is the work done by the ICM as it expands.  
$\epsilon_{\rm DM}|E_{\rm DM}|$ is the energy transferred to the ICM during major halo mergers via dynamical friction. Note that the exact value of $\epsilon_{\rm DM}$ remains highly uncertain and depends on other factors such as the merger history of a given halo.  We use the outer accretion shock as the boundary of the gas, which is approximately $1.89$ times the splashback radius $R_{\rm sp}$ informed by results from cosmological simulations \citep{aung_etal21}. For the value of $R_{\rm sp}$, we use the fitting function from Equation~(7) in \citet{more_etal15} which relates $R_{\rm sp}$ with the halo peak height $\nu_{200m} \equiv 1.686/\sigma(M_{200m},z)$, where $\sigma(M,z)$ is the variance of density fluctuations at the redshift $z$ on the mass scale $M$.
This sets the boundary condition for solving the momentum conservation equation. 

The term $\epsilon_f M_\star c^2$ represents the energy injected into the ICM due to feedback from both supernovae (SNe) and active galactic nuclei (AGN), where $M_\star$ is the total stellar mass. The stellar mass is given by the stellar fraction $F_\star(<R_{500c}) = M_\star(<R_{500c})/M_{500c}$, which depends only on the halo mass as
\begin{equation}\label{eq:fstar}
F_\star (M_{500c}) = f_\star\left(\frac{M_{500c}}{3\times10^{14}M_\odot}\right)^{-S_\star}.
\end{equation}
which is described by two parameters $(f_\star, S_\star)$ that control the normalization and the slope of the $F_\star-M$ relation. Following \citep{flender_etal17}, we choose $f_\star = 0.026$ and $S_\star = 0.12$ as our fiducial parameters. These values are chosen to match the values of the observed stellar mass fraction \citep[see Table~2 in][]{flender_etal17}.

Alternatively, the gas mass fraction can be set instead of the stellar mass fraction:
\begin{equation}\label{eq:fgas}
F_{\rm gas} (M_{500c}) = f_{\rm gas}\left(\frac{M_{500c}}{3\times10^{14}M_\odot}\right)^{S_{\rm gas}}.
\end{equation}
where the two free parameters are $(f_{\rm gas}, S_{\rm gas})$. The sum of the gas mass fraction and stellar mass fraction is assumed to be equal to the cosmic baryon fraction $f_{b} = \Omega_b/\Omega_M$, independent of halo mass and redshift:
\begin{equation}\label{eq:fbaryons}
  F_{\rm gas} (M_{500c}) +   F_\star (M_{500c}) = f_{b}. 
\end{equation}

We use a radially-dependent non-thermal pressure from \citet{nelson_etal14b}:  
\begin{equation}
f_{\rm nt} \equiv \frac{P_{\rm rand}(r)}{P_{\rm tot}(r)} =  1 - A_{\rm nt}\left[1+\exp\left\{-\left(\frac{r}{B_{\rm nt}\, R_{ 200m}}\right)^{C_{\rm nt}}\right\}\right], \label{eq:nelson}
\end{equation}
where $R_{\rm 200m}$ is the spherical over-density radius with respect to 200 times the mean matter density of the universe. The parameters are calibrated to be $A_{\rm nt} = 0.452$, $B_{\rm nt} = 0.841$, $C_{\rm nt} = 1.628$. The $R_{\rm 200m}$ scaling ensures halo redshift and mass independence at the cluster scales $M_{500c} \geq 3\times 10^{14} h^{-1}M_\odot$ over which this relation is calibrated with the {\em Omega500} cosmological simulation \citep{nelson_etal14b}. The thermal pressure $P_{\rm th}$ is obtained by multiplying $P_{\rm tot}$ with $(1-f_{\rm nt})$. 
Note that in our model, instead of modeling the thermal pressure $P_{\rm th}$ directly, $P_{\rm th}$ is derived from the total pressure $P_{\rm tot}$ and the non-thermal pressure fraction $f_{\rm nt}$. Physically, this is because $P_{\rm tot}$ can be better  described by a polytropic equation of state than that of $P_{\rm th}$. Specifically, $P_{\rm tot}$ results in a constant polytropic index across a wide range of gas density, while $P_{\rm th}$ requires a density-dependent $\Gamma$ (see Figure~1 in \citealt{shaw_etal10}). 

Table~\ref{tab:param} summarizes the parameters of the BP gas model, their physical meanings, and default values.  The BP model is flexible for describing gas profiles in cosmological simulations and observations. In Appendix~\ref{sec:model_accuracy}, we show the comparison of the density and pressure profiles between the BP model and those of the IllustrisTNG300 simulations and observations.

\subsection{Calculating X-ray and thermal SZ Observables}

We compute the X-ray emissivity profile of a given halo as
\begin{align}\label{eq:xray_emissivity}
\epsilon_X (r; M_{\rm vir},z) &= \frac{n_H(r; M_{\rm vir},z) n_e(r; M_{\rm vir},z)}{4\pi(1+z)^4} \\ \nonumber
&\times \int^{E_{\rm max}(1+z)}_{E_{\rm min}(1+z)}\Lambda(T(r; M_{\rm vir},z),Z,E) dE
\end{align}
where $n_H$, $n_e$ and $T$ are the hydrogen and electron number densities and gas temperature respectively. We use the {\tt APEC} plasma code version 3.0.9 \citep{foster_etal12} to compute the X-ray cooling function $\Lambda$, integrated over the energy range $[E_{\rm min}, E_{\rm max}]$ in the observer's frame. The fiducial values for $E_{\rm min}$ and $E_{\rm max}$ are $0.5$ and $2.0$~keV respectively. We assumed a constant metallicity of $Z=0.3 Z_\odot$ throughout the ICM, as suggested from observations \citep[e.g.,][]{mernier_etal18}, and we used the Solar abundance values from \citet{asplund_etal09}.  

The Compton-$y$ profile of a given halo is directly derived from the pressure profile of thermal electrons:
\begin{equation}\label{eq:compton-y}
    y(r; M_{\rm vir},z) = \frac{\sigma_T}{m_e c^2} \frac{\mu}{\mu_e} P_{\rm th}(r; M_{\rm vir}, z),
\end{equation}
where $\sigma_T = 6.25 \times 10^{-25}\,{\rm cm^2}$ is the Thomson cross-section, $m_e c^2 = 511\,{\rm keV}$ is the electron rest mass energy, and $P_{\rm th}(r;M_{\rm vir},z)$ is the thermal gas pressure profile. Here $\mu = 4/(5X + 3)$ and $\mu_e = 2/(X + 1)$ are the mean molecular weights of the fully ionized gas and electron, respectively, with $X = 0.76$ the primordial hydrogen mass fraction. 

\subsection{Triaxial Gas Halo and its 2D Projection}\label{sec:triaxial_model}

\begin{figure}
    \centering
    \includegraphics[width=1.0\columnwidth]{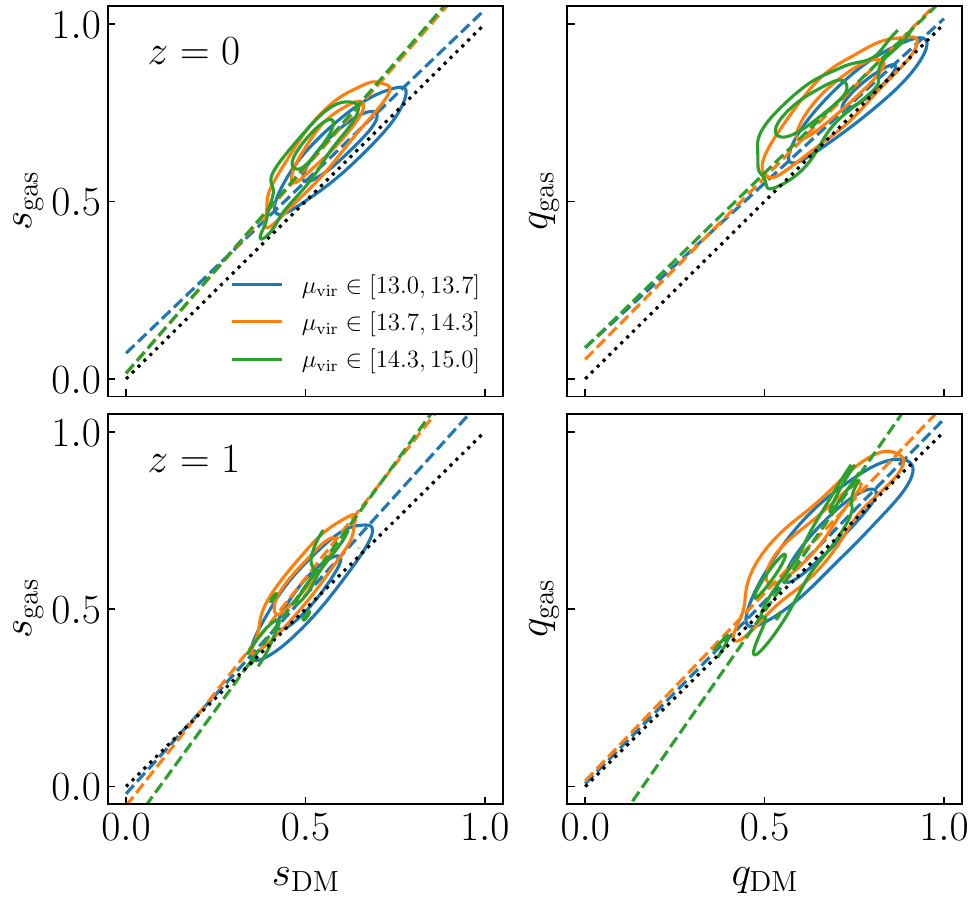}
    \caption{
    Plots of gas shape versus DM shape in halos in different mass bins with $\mu_{\rm vir} \equiv \log_{10} (M_{\rm vir}/M_\odot)$ from the TNG300 simulations at $z=0$ (upper panels) and $z=1$ (lower panels). The solid lines indicate the $1,2,3\sigma$ of the density distributions of the number of halos. The dashed lines are the bisector fits to the relation between gas and DM shape. The {\em left} panel shows the short-to-long axis ratio $s=c/a$, and the {\em right} panel shows the mid-to-long axis ratio $q=b/a$, where $a\geq b\geq c$ being the major, middle, and minor axes of the triaxial ellipsoid, respectively.  The black dotted lines show where $s_{\rm gas} = s_{\rm DM}$ and $q_{\rm gas} = q_{\rm DM}$.     
    }
    \label{fig:shape}
\end{figure}

DM and gas in halos are triaxial instead of spherical, as shown in both observations and cosmological simulations. Halo triaxiality is a natural consequence of cold DM structure formation. It is expected that halo triaxiality leads to bias and scatter in observable-mass scaling relations and bias in cluster selection functions.  However, in most analytical models or observational analyses, both the DM and gas halos are treated as spherical halos.  

In BP, we model the triaxial shapes of both DM and gas halos to investigate their impact on cluster observables. 
The BP code uses the provided axis ratios from the DM halo catalog. 

Cosmological simulations show that gas halos are more spherical than their DM hosts. To model the gas shape, we calibrate an empirical model of the dependence of the gas shape on the DM shape using the shape measurements from the IllustrisTNG300 cosmological hydrodynamical simulations. 

The gas axis ratios for the given DM axis ratios with masses between $\log_{10} (M_{\rm vir}/M_\odot) = [13.0, 15.0]$ for $0\leq z \leq 2$ are expressed as 
\begin{eqnarray}\label{eq:s_and_q}
s_{\rm gas} &=&  s_0 + s_1 s_{\rm DM}, \\
q_{\rm gas} &=&  q_0 + q_1 q_{\rm DM}. 
\end{eqnarray}
We model their mass and redshift dependence  with 
\begin{eqnarray}\label{eq:gas_shape}
s_0 &=& 1.78-1.82\mu_{14} + 0.185 \zeta, \\
s_1 &=& -3.41-4.66\mu_{14} + 0.281 \zeta, \\
q_0 &=& 0.73-0.687\mu_{14} -0.204 \zeta, \\
q_1 &=& -0.45+149\mu_{14} + 0.204 \zeta, 
\end{eqnarray}
where $\mu_{14} \equiv \log_{10} (M_{\rm vir}/ 10^{14} M_\odot)$ and $\zeta \equiv \log_{10} (1+z)$. Figure~\ref{fig:shape} shows the relations between DM and gas triaxial ratios and their best-fits. Note that these relations are calibrated on the TNG300 simulations. In principle, they depend on the underlying baryon physics \citep{kazantzidis_etal04}, which varies between different simulations. 

We extend the analytical procedure to transform a spherically symmetric radial profile into a triaxial one by extending the formalism from \citet{stark77}. 
Denoting $(x,y,z)$ and $(x',y',z')$ as the coordinate frames of the triaxial halo and the observers respectively, where $z'$ aligns with the line-of-sight (LOS) direction, the two reference frames are related by the transformation $(x,y,z)^T = \mathbf{A}(x',y',z')^T$
where
\begin{equation}
\mathbf{A}=
\begin{pmatrix}\label{eq:rot_mat}
    c_1 c_3 -s_1c_2s_3 & -c_1 s_3 - s_1 c_2 c_3 & s_1 s_2 & \\
    s_1 c_3 + c_1 c_2 s_3 & -s_1 s_3 + c_1 c_2 c_3 & -c_1 s_2 \\
    s_2 s_3 & s_2 c_3 & c_2
\end{pmatrix}.
\end{equation}
Here we adopt a short-hand notation for $\cos$ and $\sin$ as $c$ and $s$ respectively, and $(1,2,3)$ stands for the Euler angles $(\phi,\theta,\psi)$ that specify the rotation. Specifically, $\theta$ is the angle between the major axis with the line-of-sight axis $z'$, with $\theta \in [0,\pi/2]$, while $\phi \in [0,2\pi)$ and $\psi \in [0,2\pi)$. 

For any generic 3D spherical profile $f_{3D}(r)$, we can substitute the spherical radius $r$ with the ellipsoidal radius $r_{\rm ep}$, where the triaxial profile is then described by $f_{3D}(r_{\rm ep})$. The ellipsoidal radius in the observer's frame (the primed frame) is defined as
\begin{equation}
r_{\rm ep}^2 = f z'^2 + gz' + h,
\end{equation}
where
\begin{align}
f &= A_{13}^2/s^2 + A_{23}^2/q^2 + A_{33}^2, \\
g &= 2\left(
A_{11}A_{13}/(s^2) + A_{21}A_{23}/(q^2)+ A_{31}A_{33} \right) x' \\ \nonumber
&+ 2\left( A_{12}A_{13}/(s^2) + A_{22}A_{23}/(q^2) + A_{32}A_{33} \right) y',\\
h &= \left(A_{11}^2x'^2 + A_{12}^2y'^2 + 2A_{11}A_{12}x'y' \right)/s^2 \\  \nonumber
&+ \left(A_{21}^2x'^2 + A_{22}^2y'^2 + 2A_{21}A_{22}x'y' \right)/q^2 \\  \nonumber
&+ \left(A_{31}^2x'^2 + A_{32}^2y'^2 + 2A_{31}A_{32}x'y' \right), 
\end{align}
where $A_{ij}$ are the entries of the rotational matrix $\mathbf{A}$ in Equation~(\ref{eq:rot_mat}), $q = b/a$, $s=c/a$ are the axis ratios, with $a\geq b\geq c$ being the major, middle, and minor axes of the triaxial ellipsoid respectively. 

We then project the 3D triaxial profile $f_{3D}(r_{\rm ep})$ to the 2D distribution given by
\begin{align}
   F_{2D}(x',y') &\equiv \int_{-r_{\rm max}}^{r_{\rm max}} f_{3D}(r_{\rm ep}) dz' \\ \nonumber 
   &= 2\sqrt{f}\int_{0}^{r_{\rm max}} f_{3D}(\sqrt{z_\star^2 + \zeta^2}) dz',
\end{align}
where $z_\star= \sqrt{f} \left( z' + g/(2f)\right)$, $\zeta = h - {g^2}/{(4f)}$,
and we set $r_{\rm max} = \sqrt{f} \left( 5R_{500c} + g/(2f)\right)$, corresponding to the outer accretion shock radius of the gaseous halo at $\sim  5R_{500c}$ \citep{aung_etal21}. 

\subsection{Modeling intrinsic scatter in halo gas profiles}\label{sec:intrinsic_scatter}

\begin{figure}
    \centering
    \includegraphics[width=0.8\columnwidth]{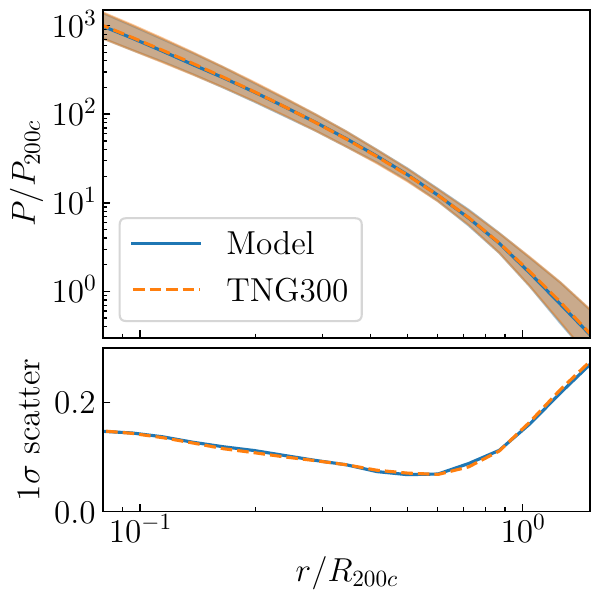}
    \includegraphics[width=0.8\columnwidth]{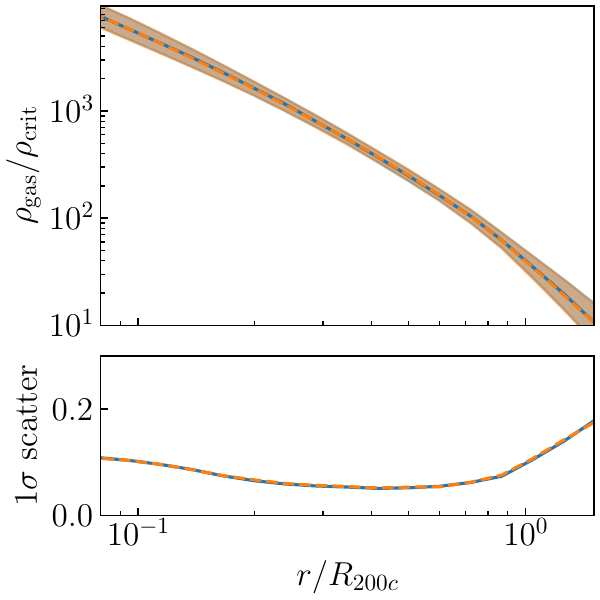}
    \caption{Top and bottom panels show the model pressure and density profiles sampled from the covariance matrix computed from the TNG300 simulation (solid blue), compared to the input TNG300 profiles (dashed orange). The lines and the shaded regions are the mean profiles $1\sigma$ scatter (over the natural logarithm of the profiles). The bottom subpanels in both panels show the $1\sigma$ scatter for the sampled and input TNG profiles. 
    }
    \label{fig:pgas_cov}
\end{figure}

BP map-making also includes a method of incorporating variations in the gas profiles due to differences in their formation histories and baryonic physics. Specifically, we adopt a non-parametric, empirical approach using the covariance of gas profiles measured from empirical data or cosmological simulations, following \citet{comparat_etal20}. 

In the model presented here, we use the IllustrisTNG300 simulations to compute the covariance matrices for the logarithm of the thermal pressure and gas density. 
Specifically, for a generic profile $f(r)$ for each mass and redshift bin, we measure the covariance matrix $\mathcal{C}(r,r')$ as
\begin{equation}
    \mathcal{C}(r,r') = \langle (\ln f(r)-\overline{\ln f}(r))(\ln f(r')-\overline{\ln f}(r')) \rangle
\end{equation}
summing over all halos in the bin, where $\overline{\ln f}(r)$ is the mean profile in natural logarithm at radius $r$. 
We normalize the gas profiles with respect to their self-similar quantities:
\begin{eqnarray}
    P_{200c} &=& \frac{GM_{200c}}{2R_{200c}} 200 \rho_{\rm crit}(z) f_b, \\
    \rho_{200c} &=& 200 \times \rho_{\rm crit}(z),
\end{eqnarray}
where $f_b = \Omega_b/\Omega_M$ is the cosmic baryon fraction, and $\rho_{\rm crit}(z)$ is the critical density of the Universe at redshift $z$. We account for additional halo mass dependence in the pressure profiles due to baryonic physics by applying the Kernel Localized Linear Regression (KLLR) method \citep{kllr} to estimate the average halo mass trend of the pressure profile in each scaled radial bin in $r/R_{200c}$, where a Gaussian kernel is applied to get the average pressure as a function of $M_{200c}$. 

We then generate the model variation profile $\delta \ln f(r)$ by sampling the covariance matrix, treating the covariance matrix as a multivariate Gaussian distribution. For $N=25$ radial bins, the multivariate Gaussian distribution is 
\begin{eqnarray}
    &&P(\delta \ln f_1, \delta \ln f_2, \dots, \delta 
    \ln f_N) \\ \nonumber &&=\frac{1}{\sqrt{(2\pi)^N|\mathcal{C}|}}\exp\left(-\frac{1}{2}({\delta \ln f})\mathcal{C}^{-1}({\delta \ln f} )^T\right),
    \label{eqn:log_normal}
\end{eqnarray}
where ${\delta \ln f} = (\delta \ln f_1, \delta \ln f_2, \dots, \delta \ln f_N)$ is a random variable that represents the deviation from the mean log profile with mean $\overline{\delta \ln f} = 0$. Once the covariance matrix $\mathcal{C}$ is given, we can draw a realization of the variation in the profile $\delta \ln f$ from the multivariate Gaussian distribution. The resulting realization of the profile is then the sum of the mean profile and the variation: $ \ln f = \overline{\ln f} + \delta \ln f$. 

Figure~\ref{fig:pgas_cov} shows the normalized thermal pressure and density profiles sampled from the covariance matrices of pressure and density measured from the TNG300 simulations, compared to the profiles directly measured from the TNG300 simulations. It shows the covariance sampled profiles recover 1$\sigma$ scatter of the simulation profiles. 

Note that the profile covariance is model-dependent, as it is derived directly from measurements on the cosmological simulation we use, in this case TNG300. The profile covariance matrices can be different if we derive them from another set of simulations. 

\begin{figure*}
    \centering	\includegraphics[width=1.\columnwidth]{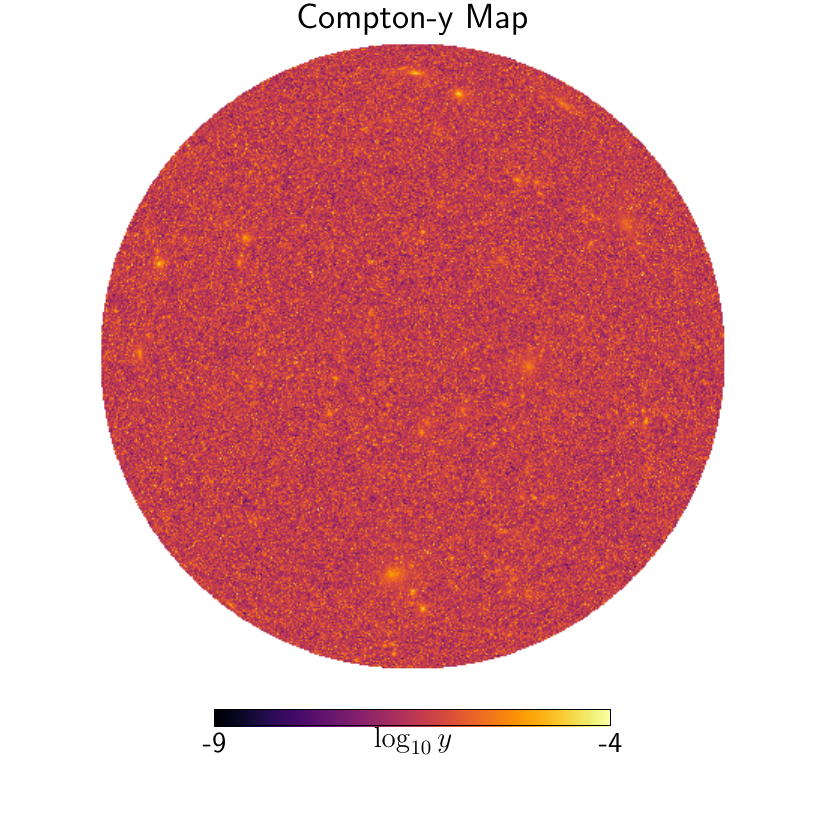}
    \includegraphics[width=1.\columnwidth]{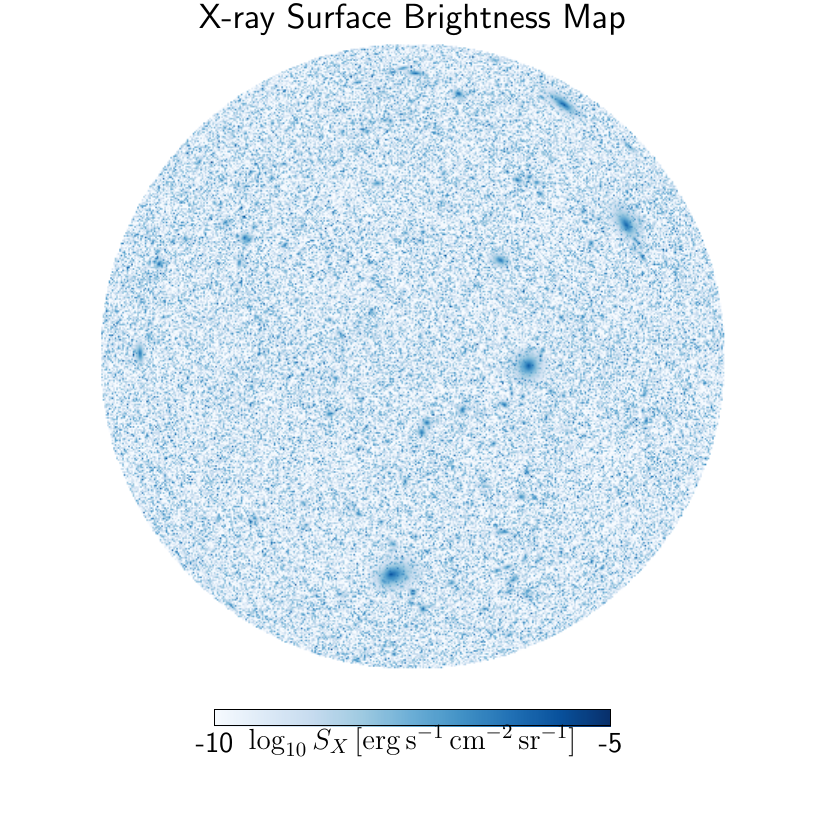}
    \caption{ Half-sky Compton-$y$ map ({\em left} panel) and X-ray surface brightness map {\em right} panel) generated by applying the BP model to the Uchuu Lightcone. These maps are generated with triaxial halos but no intrinsic scatter. See Section~\ref{sec:bp_maps} for more details. }
    \label{fig:uchuu_maps}
\end{figure*}

\section{Map Making}\label{sec:bp_maps}

\subsection{Uchuu Lightcone}
The Uchuu Lightcone covers half of the sky from $0<z<2$. The lightcone is based on the Uchuu Simulation, a large-scale DM-only cosmological simulation with a box size of $(2\,h^{-1}{\rm Gpc})^3$ with DM particle resolution of $m_p = 3.27\times 10^8\,h^{-1}M_\odot$ \citep{uchuu}. The simulation is performed assuming {\em Planck} cosmology with $\Omega_{m}
= 0.3089, \Omega_b = 0.0486, n_s=0.9667, h=0.6774, \sigma_8=0.8159$, and $w=-1.0$. For the rest of the paper, we adopt the same cosmology unless stated otherwise.  
Halos and subhalos are identified by the ROCKSTAR code~\citep{Behroozi2013}.\footnote{\url{https://bitbucket.org/gfcstanford/rockstar/}}
Other Uchuu data products, such as mock galaxy catalogs based on Uchuu-UniverseMachine \citep{Aung2023, Prada2023}, Uchuu-$\nu^2$GC \citep{Oogi2023}, Uchuu-SDSS \citep{Paez2024,Garcia2024}, GLAM-Uchuu Lightcone \citep{Ereza2024}, and infrared sky SIDES-Uchuu \citep{Gkogkou2023}, are publicly available in the \textit{Skies $\&$ Universes} database.\footnote{\url{http://www.skiesanduniverses.org/Simulations/Uchuu/}}

We used 27 snapshots between $z=0$ and $2$ of the Uchuu simulation to construct the half-sky Uchuu Lightcone.  We place an observer in the simulation box and then transform the Cartesian coordinates of each halo into equatorial coordinates.  The redshift of each halo is calculated by using the line-of-sight distance.  We select halos between the redshift $(z_{i-1} + z_i)/2$ and $(z_i + z_{i+1})/2$ in the given snapshot $i$, where $z_i$ is the redshift of this snapshot.  Because the Uchuu $(2\,h^{-1}{\rm Gpc})^3$ volume is not enough to cover the half-sky spherical volume at a higher redshift, box replications are necessary.  Instead of periodic replication, we apply three randomization transformations for each replica: the box rotation, mirroring, and translation \citep{Blaizot2005, Bernyk2016}, and then tile them to cover the spherical shell at a given snapshot.  When the center of a halo lies close to the edge of a given spherical shell, it sometimes happens that the parts of subhalos of this halo do not lie within the shell. To ensure the hierarchy of halo and subhalo, we include such subhalos in the given shell. Finally, we join all spherical shells together to construct the half-sky lightcone. 

The Uchuu Lightcone contains halos with a mass range of $M_{500c} \geq 10^{13} M_\odot$ and a redshift range of $0\leq z \leq 2$, with a total of $75,159,192$ halos.  The lightcone catalog also contains information on halo concentrations, ratios of the halo axis $c/a$ and $b/a$, and the orientation of the major axis, derived from the ROCKSTAR halo catalog. 

\subsection{Generation of the Baryon Pasted Maps}

We generate the maps in X-ray surface brightness in energy bands of $0.5-2$~keV and the tSZ Compton-$y$ maps in HEALPix (Hierarchical Equal Area and iso-Latitude Pixelization) \citep{healpix} projection, with $N_{\rm side} = 8192$, corresponding to angular pixel size of about $25.7$ arcseconds. 
We generate the maps by taking the RA, Dec, mass, and redshift provided in the Uchuu Lightcone catalog. We then map the RA and Dec positions of the 2D halo profile $\Sigma_{2D}(x',y')$ by determining which HEALPix pixels the profile belongs to, using the {\tt query\_disc\_inclusive} and {\tt pix2ang} functions provided by the HEALPix C++ package.\footnote{\url{https://healpix.sourceforge.io/html/Healpix_cxx/index.html}}

To study the impact of triaxiality and intrinsic scatter in X-ray and tSZ observables, we generated different realizations of the Compton-$y$ and X-ray surface brightness maps with the Uchuu Lightcone: 
\begin{itemize}
    \item spherical halos without intrinsic scatter (sph),
    \item triaxial halos with no intrinsic scatter (tri),
    \item spherical halos with intrinsic scatter (sph+var),
    \item triaxial halos with intrinsic scatter (tri+var),
\end{itemize}  
with a total of 8 maps (4 X-ray Surface Brightness, 4 Compton-$y$). Details of the triaxial halo projection and intrinsic scatter modeling can be found in Sections~\ref{sec:triaxial_model} and \ref{sec:intrinsic_scatter}, respectively. Note that no foreground or background noise is applied to these maps. Figure~\ref{fig:uchuu_maps} shows the X-ray and Compton-$y$ maps with triaxiality but no intrinsic scatter (the `tri' maps) in the gas profiles. 
Note that maps with triaxial halos with intrinsic scatter (`tri+var') overestimate the level of total scatter, since intrinsic scatter that we derived empirically from cosmological simulations does include the contribution from triaxiality. The scatter derived from this map thus provides an upper limit to the total amount of scatter expected from both intrinsic scatter and triaxiality.
Note that for these sets of maps, we do not use the halo concentration provided by ROCKSTAR to model the gas profiles. This is because halo concentration can contribute to extra scatter in the profile, and is also correlated with halo triaxiality \citep{lau_etal2020}. To avoid double-counting the scatter, we fix the halo concentration to be the same for halos with the same mass and redshift using the fitting function from \citet{diemer_kravtsov15}.

\begin{figure*}
    \centering
    \includegraphics[width=2.0\columnwidth]{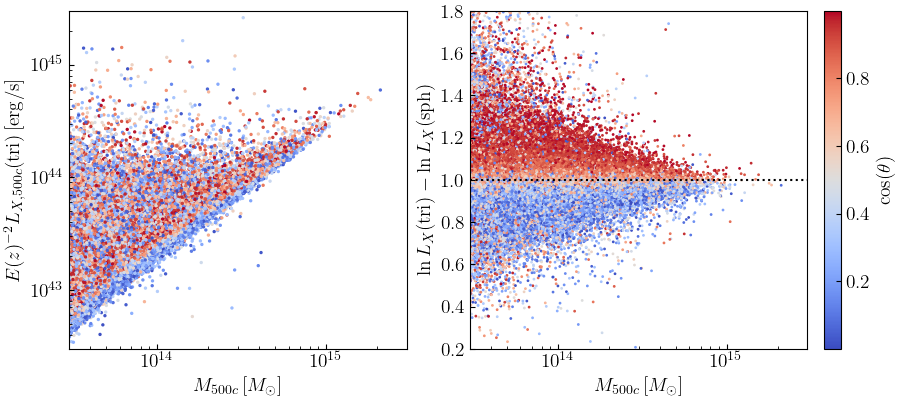}
    \caption{The {\it Left} panel shows X-ray luminosity-Mass scaling relation with the X-ray luminosity $L_X$ extracted from circular apertures with radii $R_{500c}$ in cluster-size halos $M_{500c} \geq 5\times 10^{13} M_\odot$ at $z<0.5$ from the X-ray Surface Brightness Uchuu BP Map, generated with halo triaxiality (the `tri' map). The $E(z)^{-2} = H(z=0)/H(z)$ factor accounts for the self-similar redshift evolution of $L_X$.  The color indicates the magnitude of $\cos(\theta)$, where $\theta$ is the angle between the major axis of the halo and the line-of-sight. Halos with high values of $\cos(\theta)$, i.e. halos with major axes more aligned with the line-of-sight, drives up the scatter in $L_X$. This is more evident in the {\it right} pane where we show the ratio between the $L_X$ in the `tri' map, to that extracted from the spherical `sph' map. Triaxial halos with higher (lower) $\cos(\theta)$ shows larger (smaller) $L_X$ values compared to their spherical counterparts.  
    }
    \label{fig:diff}
\end{figure*}

\begin{figure}
    \centering
    \includegraphics[width=1.0\columnwidth]{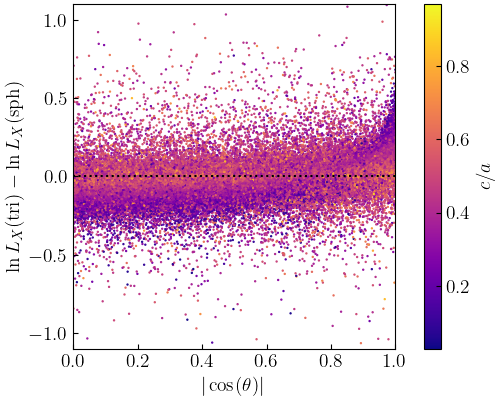}
    \caption{This plot shows the ratio between the $L_X$ extracted from the halos in the triaxial BP map (`tri'), to that extracted from the spherical BP map (`sph'), as a function of the orientation magnitude $\cos(\theta)$. The color indicates the minor-to-major axis ratio $c/a$. It shows that the scatter in $L_X$ due to triaxiality is driven mostly by elongated halos with $c/a \lesssim 0.5$. They have lower projected $L_X$ when their major axis is lying closer to the plane of sky (e.g., $|\cos\theta| < 0.8$ ) and drive the projected $L_X$ values high when their major axes are more aligned with the line-of-sight (e.g., $|\cos\theta| > 0.8$). 
    }
    \label{fig:Lx_s_orientation}
\end{figure}

\begin{figure}
    \centering
    \includegraphics[width=1.0\columnwidth]{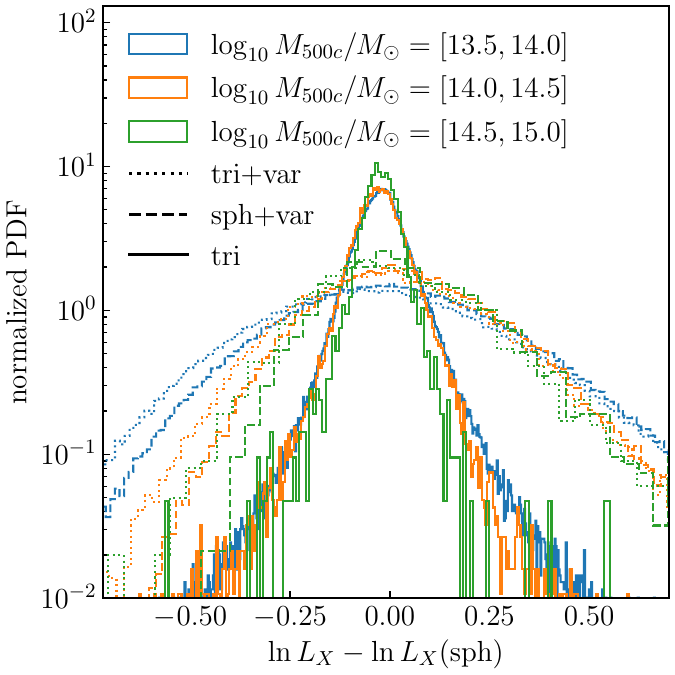}
    \caption{This plot shows the distribution of the $L_X$ extracted from the halos in the triaxial XSB map (`tri', solid lines), in the XSB map with spherical halos plus intrinsic scatter (`sph+var', dashed lines), and XSB map with triaxial halos with intrinsic scatter (`tri+var', dotted lines), to that extracted from the spherical BP XSB map (`sph'), in different halo mass bins. 
    It shows that the scatter due to triaxiality is smaller than the intrinsic scatter. Lower mass halos have larger intrinsic scatter and larger triaxial scatter, compared to higher mass halos. 
    }
    \label{fig:Lx_scatter_dist}
\end{figure}

\subsection{Performance, Memory and Storage Requirements}

The BP map-making code is implemented in C++ and parallelized using Message Passing Interface (MPI), making it fast and efficient to generate maps with a large number of halos. The map-making approach presented in this paper is based on halo-by-halo, which is much faster than the particle-based approach presented in \citet{bp_algo}, especially since each halo comprises at least 1000 DM particles.  The code allocates halos across MPI tasks. Given the substantial number of halos ($75$ million) in the Uchuu lightcone, it is necessary to partition the lightcone into smaller parts to accommodate the data within the computer's memory. We partitioned the Uchuu Lightcone into 40 redshift slices and generated maps separately for each slice independently, later combining them into one.

When we generated the map, we had to balance the number of cores allotted to each MPI task with the memory accessible per task. Allocating too many MPI tasks to a fixed number of nodes depletes the memory available for each task, while too few MPI tasks slow down the map-making process. On Yale's {\tt Grace} machine, featuring 48 cores (Intel XEON Icelake 2.40~GHz) and 480~GB of RAM per node, we allocated 16 MPI tasks across 16 nodes (one task per node). Consequently, projecting a single halo took about 2 minutes of wall time. This performance could be improved by assigning more MPI tasks per map with additional nodes. Regarding storage needs, for the BP-Uchuu map, the number of pixels is $N_p = 12N_{\rm side}^2 \approx 8 \times 10^8$ with $N_{\rm side}=8192$, resulting in a map file size of approximately $6$~GB.

\section{Results}\label{sec:science}

\subsection{Extrinsic Scatter in X-ray Luminosity--Halo Mass Scaling Relation}

Figure~\ref{fig:diff} shows the impact of the triaxiality of the halo on the integrated X-ray luminosity $L_X$. The left panel shows the X--ray luminosity--mass scaling relation in the `tri' XSB map. $L_X$ is measured within a circular aperture of radius $R_{500c}$ for each halo. The halos considered here have masses $M_{500c} \geq 5 \times 10^{13} M_\odot$ and redshifts $z \leq 0.5$. The color of each data point represents the alignment of the triaxial halo with the line-of-sight, quantified by $\cos \theta$, where $\theta$ is the angle between the major axis of the halo and the line-of-sight. For a given mass, halos with lower values of $|\cos \theta | \lesssim 0.2$ generally exhibit lower values of $L_X$. 

In the right panel, we show the comparison in $L_X$ between halos in the `tri' and `sph' maps. Each data point represents the difference in $\ln L_x$ measured from the `tri' map to that from the `sph' map for the same halo. This allows us to factor out the dependence on mass, as well as projection due to other halos (i.e., contamination from 2-halo terms), since the same halo on both maps is subject to the same projection. Halos with high $|\cos\theta|$ have $\ln L_X$ biased higher than those with low $|\cos\theta|$, and vice versa. 

Figure~\ref{fig:Lx_s_orientation} shows how the differences in $\ln L_X$ between `tri' and `sph' halos depend on $\cos\theta$ and the minor-to-major axis ratio $c/a$. It shows that the scatter due to triaxiality is driven mostly by elongated halos with low values of $c/a \lesssim 0.5$, dependent on their orientations: halos drive the bias high when their major axes are more aligned with the line-of-sight for $|\cos\theta| > 0.8$, and the biases are lower when their major axes lie nearer to the plane of the sky for $|\cos\theta| < 0.8$. 

Figure~\ref{fig:Lx_scatter_dist} shows the distribution of $\Delta \ln L_X$ due to triaxiality, intrinsic scatter, and triaxiality plus intrinsic scatter from the `tri', `sph+var', and `tri+var' maps, respectively. Using the $1\sigma$ standard deviation in $\Delta \ln L_X$ as the proxy for scatter, the cluster-sized halos with $\log_{10}(M_{500c}/M_\odot) \in [14.5, 15.0]$ have a scatter of $8\%$, which is comparable to the $9\%$ scatter in groups with $\log_{10}(M_{500c}/M_\odot) \in [13.5, 14.0]$ and less massive clusters with $\log_{10}(M_{500c}/M_\odot) \in [14.0, 14.5]$. The scatter due to triaxiality is subdominant to the intrinsic scatter, shown in dashed lines in the same figure. The intrinsic scatter is dependent on halo mass, peaking at $29\%$ for group-size halos and dropping to $20\%-22\%$ for cluster-size halos. When combining triaxiality and intrinsic scatter, the total scatter reaches $30\%$ for group-sized halos and $20\%$ for clusters.

Excluding halo core regions can significantly reduce intrinsic scatter in X-ray luminosity. By omitting pixels within a circular aperture of $0.2 R_{500c}$ around each halo's center, the intrinsic scatter decreases from $29\%$ to $22\%$ for groups and from $20\%$ to $11\%$ for massive clusters. However, the scatter due to triaxiality remains unchanged after core excision, maintaining approximately $12\%$ at the group scale and $7\%$ at the cluster scale, which constitutes nearly half of the total scatter. These results indicate that the halo outskirts are the primary contributors to triaxial scatter. Our results are qualitatively similar to the previous work based on hydrodynamical cosmological simulations \citep[e.g.,][]{battaglia_etal12}.

Table~\ref{tab:Lx_scatter} summarizes the scatter for all combinations of triaxiality and intrinsic scatter, with and without core excision. 

\begin{table}
\begin{center}
    \begin{tabular}{|c|c|c|c|}
        \hline
         & \multicolumn{3}{c|}{$\log_{10}(M_{500c}/M_\odot)$}\\
         \cline{2-4}
         map type & $[13.5, 14.0]$ & $[14.0, 14.5]$ & $[14.5, 15.0]$ \\ 
        \hline
        \multicolumn{4}{|c|}{Without core excision} \\       
        \hline
        tri  & 0.09 & 0.08 & 0.08  \\ 
        sph+var & 0.29 & 0.22 & 0.20 \\
        tri+var & 0.31 & 0.24 & 0.22\\
        \hline
        \multicolumn{4}{|c|}{With core excision} \\       
        \hline
        tri  & 0.12 & 0.09 & 0.07  \\ 
        sph+var & 0.22 & 0.15 & 0.11 \\
        tri+var & 0.26 & 0.19 & 0.15\\
        \hline
    \end{tabular}
    \caption{$1\sigma$ scatter in $\ln L_X({\rm map\, type})- \ln L_X({\rm sph})$ for halos $z<0.5$ in different halo mass bins, with and without the core ($r<0.2 R_{500c}$) excised. } 
    \label{tab:Lx_scatter}
\end{center}
\end{table}

\subsection{Bias in X-ray and Thermal SZ Power Spectra due to Intrinsic Scatter in ICM Profiles}

\begin{figure*}[htbp]
\centering
    \includegraphics[width=2\columnwidth]{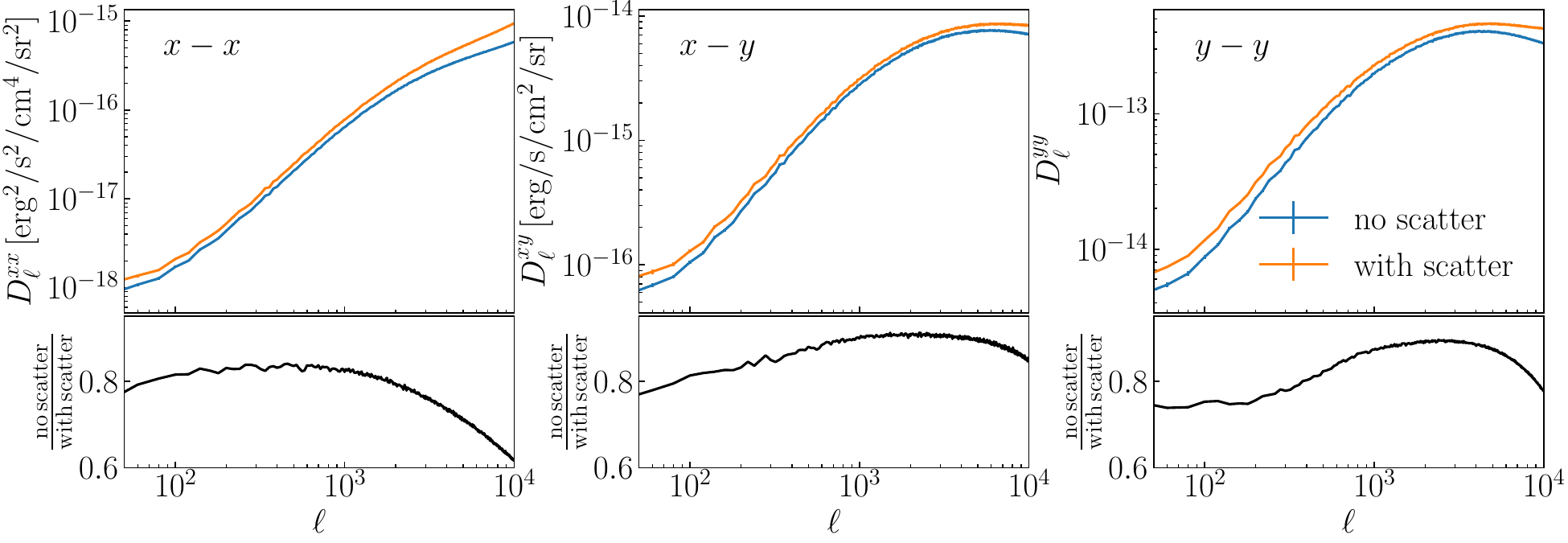}
    \caption{Plots showing the effect of intrinsic scatter of pressure and density profiles on the X-ray and tSZ auto- and cross-angular power spectra. The left, middle, and right panels show the X-ray auto-power spectra, X-ray/tSZ cross-power spectra, and tSZ auto-power spectra,  measured from the Uchuu BP maps respectively. In each panel, the blue and orange lines represent the power spectra with and without intrinsic scatter in the halo profiles. The bottom sub-panels show the ratios of the power spectra with intrinsic scatter in the halo profiles, to those without. Including intrinsic scatter in the gas profiles increases the normalizations of the power spectra. Note that the errorbars in each of the power spectrum are much smaller than the difference between the spectra.  
    Also note that the tSZ auto-power spectrum is unit-less. 
    }
    \label{fig:scatter_power_spectra}
\end{figure*} 

Figure~\ref{fig:scatter_power_spectra} illustrates the X-ray and tSZ auto- and cross-angular power spectra derived from the `sph' map (spherical halos with no intrinsic scatter) and the `sph+var' map (spherical halos with intrinsic scatter). These spectra were computed with the {\tt anafast} routine in {\tt healpy}. We masked out $z<0.1$ halos to reduce non-Gaussian cosmic variance. The results demonstrate that intrinsic scatter in the density and pressure profiles significantly increases the power of both the X-ray and the tSZ angular power spectra in a scale-dependent way. Specifically, intrinsic scatter leads to = $20\%$ increase in the X-ray auto-power spectrum at small multipoles $\ell = 100 $, which increases to $40\%$ at $\ell = 10^4$. The intrinsic scatter increases the tSZ auto-power spectrum by $25\%$ at $\ell = 100$, by $10\%$ at $\ell = 3000$, and by $20\%$ at $\ell = 10^4$.  
The X-ray/tSZ cross-power spectrum shows a modest increase by $25\%$ at $\ell = 100$, by $10\%$ at $\ell = 3000$, and by $15\%$ at $\ell = 10^4$. The contribution to the increase in power is dominated by massive halos $\log_{10}( M_{\rm vir}/M_\odot) \geq 14.5$ at low redshift $z < 0.5$.
This rise in power due to intrinsic scatter in the gas profiles could potentially bias cosmological parameter estimates, particularly $\sigma_8$, which is sensitive to the normalization of the angular power spectrum. Using the $C_\ell \propto \sigma_8^{8.1}\Omega_M^{3.2}H_0^{-1.7}$ scaling \citep{bolliet_etal18}, a change in the amplitude of the tSZ power spectrum $\Delta C_\ell/C_\ell \sim 25\%$ at $\ell\sim 100$ translates into bias in $\Delta \sigma_8/\sigma_8 \sim 3\%$, $\Delta \Omega_M/\Omega_M \sim 8\%$, and $\Delta H_0/H_0 \sim -15\%$, if the scatter in the thermal pressure profiles is neglected. Note that halo triaxiality has no effect on the measured power spectrum. 

The observed increase in power can be interpreted as additional fluctuations resulting from halo-to-halo variations in pressure and density. To further explain this enhancement in angular power due to profile scatter, we provide a simple analytical framework using the halo model formalism.  In the halo model, the angular power spectrum $C$ at a given multipole $\ell$, for two different observables $\cal A$ and $\cal B$ for halo at some mass $M$ and redshift $z$ is given by
\begin{eqnarray}
C_{{\cal A}{\cal B}} &=& C^{\rm 1h}_{{\cal A}{\cal B}} + C^{\rm 2h}_{{\cal A}{\cal B}}, \\
C^{\rm 1h}_{{\cal A}{\cal B}}(\ell)
&=& \int {\rm d}z \frac{{\rm d}V}{{\rm d}z{\rm d}\Omega} 
\int {\rm d}M \frac{{\rm d}n}{{\rm d}M} \\ \nonumber
&\times& \left|{\cal A}_{\ell}(M,z) {\cal B}_{\ell}(M,z)\right|, \label{eq:A_B_1h} \\
C^{\rm 2h}_{{\cal A}{\cal B}}(\ell)
&=& \int {\rm d}z \frac{{\rm d}V}{{\rm d}z
{\rm d}\Omega} P_{\rm L}(k,z) \\ \nonumber
&\times& \left[
\int {\rm d}M \frac{{\rm d}n}{{\rm d}M} {\cal A}_{\ell}(M,z) b(M,z)\right] \\  \nonumber
&\times& \left[
\int {\rm d}M \frac{{\rm d}n}{{\rm d}M} {\cal B}_{\ell}(M,z) b(M,z)
\right], \label{eq:A_B_2h}
\end{eqnarray}
where $P_{\rm L}(k,z)$ is the linear matter power spectrum, ${\rm d}n/{\rm d}M$ is the halo mass function, and $b$ is the linear halo bias, and ${\cal A}_{\ell}$ and ${\cal B}_{\ell}$ are the Fourier transforms in multipole space of the observables $\cal A$ and $\cal B$ respectively. The angular power spectrum can be separated into the one-halo term $C^{\rm 1h}$ and the two-halo term $C^{\rm 2h}$ which represents the correlation between observables $\cal A$ and $\cal B$ within single halos and between two different halos, respectively. When the two observables are the same $\cal A = B$, then the angular power spectrum is called the auto-power spectrum; otherwise, it is the cross-power spectrum. 

The angular power spectrum of clusters is dominated by the one-halo term $C^{\rm 1h}$ at most scales of interest ($\ell > 10$), so we will ignore the two-halo term. The observables ${\cal A}'$ and ${\cal B}'$ for a random halo in a given mass and redshift bin are expected to deviate from their expectation values ${\cal A}$ and ${\cal B}$ as,
\begin{eqnarray}
%{\cal A} &=& \overline{\cal A} + \delta{\cal A},\\ \nonumber
%{\cal B} &=& \overline{\cal B} + \delta{\cal B}
{\cal A}' &=& {\cal A} + \delta{\cal A},\\ \nonumber
{\cal B}' &=& {\cal B}  + \delta{\cal B},
\end{eqnarray}
where $\delta{\cal A}$ and $\delta{\cal B}$ are random variables. In the simplest halo model, where the $\delta$ terms are zero, ${\cal A}' = {\cal A}$ and ${\cal B}' = {\cal B}$, and resulting $C^{\rm 1h}$ term is the same as in Equation~(\ref{eq:A_B_1h}). However, in a more general scenario where the $\delta$ terms are non-zero, the $C^{\rm 1h}$ term becomes,
\begin{eqnarray}
C^{\rm 1h}_{{\cal A}{\cal B}}(\ell)
&& = \int {\rm d}z \frac{{\rm d}V}{{\rm d}z{\rm d}\Omega} \int {\rm d}M \frac{{\rm d}n}{{\rm d}M} \\ \nonumber
&&\times 
\left( {\cal A}^*_{\ell}(M,z) {\cal B}_{\ell}(M,z) + \langle \delta{\cal A}^*_{\ell}(M,z) \delta{\cal B}_{\ell}(M,z) \rangle \right),
\end{eqnarray}
where the angular brackets $\langle \cdots \rangle$ denotes ensemble average, and the $*$ superscript denotes the complex conjugate. 
Note that $\langle {\cal A}^*\delta{\cal B}\rangle = {\cal A}^*\langle \delta{\cal B}\rangle = 0 $ and $\langle{\cal B}\delta{\cal A}^*\rangle = {\cal B}\langle\delta{\cal A}^*\rangle = 0$. 
Provided that there is finite correlation between the variations of the two observables $\langle \delta {\cal A}^* \delta{\cal B} \rangle \neq 0$, the implications are as follows: if $\langle \delta {\cal A}^* \delta{\cal B} \rangle > 0$, this leads to an enhanced power spectrum, while $\langle \delta {\cal A}^* \delta{\cal B} \rangle < 0$ results in a reduction of power. For the auto-power spectrum, $\langle \delta {\cal A}^* \delta{\cal A} \rangle = \langle |\delta {\cal A}|^2 \rangle \geq 0$, indicating that any noticeable fluctuations in the observable will invariably increase the clustering power. In the X-ray/tSZ cross-power spectrum, we expect an increase in power, since X-ray surface brightness, which scales with the square of the gas density, correlates positively with thermal pressure contributing to the tSZ signal.

Interpretation of the power spectrum based on the halo model usually overlooks the intrinsic scatter in the halo gas profiles. As a result, the halo model approach typically underestimates the actual amplitude of the power spectrum. Therefore, the intrinsic scatter profile provides an explanation for the differences in the power spectrum between simulations and analytic halo models \citep[e.g.,][]{battaglia_etal12b}. Thus, cosmological parameters, such as $S_8$, inferred by statistical inference based on the halo model, are often overestimated to account for the additional power from intrinsic scatter.

\section{Discussion}\label{sec:discussion}

Our findings are subject to several limitations that will be investigated further in future work. Firstly, the BP feedback model encapsulates the SNe and AGN feedback into one unified parameter. This approach does not adequately capture the intricate interactions that occur between SNe and AGN feedback \citep[e.g.,][]{medlock_etal24}. Leveraging hydrodynamical cosmological simulations that incorporate varied SNe and AGN feedback physics across a broad spectrum of mass scales, such as those from CAMELS \citep{camels} and CarpoolGP \citep{carpoolgp}, could enhance the feedback models applied in our study.

Secondly, one can improve the physicality in both the intrinsic and the extrinsic scatter models employed in this work. Intrinsic scatter stems from two main sources: (1) variations in the underlying DM matter distributions due to variations in the halo's mass accretion histories (MAH) and (2) the stochastic feedback from AGN and SNe. The TNG300 simulation informs our model for intrinsic scatter, hence our model's reliance on its specific cosmology and feedback prescriptions. To generalize this model, we must investigate how the intrinsic scatter varies for a range of cosmology and galaxy formation models using the CAMELS and CARPoolGP simulations. 

Extrinsic scatter in our model is tied to the triaxiality of a halo gas, which is also calibrated using TNG300 and thus shares its limitations, notably its sensitivity to the cosmology and subgrid physics used in the simulation. Capturing the triaxial shape of a halo gas presents further challenges. The triaxial form of gas depends on that of the DM halo, and their relationship is influenced by baryonic physics \cite[e.g.,][]{kazantzidis_etal04,lau_etal11,machado_etal21}. Moreover, the triaxiality of the DM halo is also shaped by its MAH \citep[e.g.,][]{lau_etal2020}. Consequently, modeling gas triaxiality requires a two-step methodology: (1) determining DM halo triaxiality using MAHs from simulations with varying cosmologies and (2) constructing a gas triaxial model based on the DM triaxial configuration using a series of cosmological simulations involving diverse baryonic physics. 
As weak lensing mass is also subjected to orientation bias \citep{becker_kravtsov11}, this DM-gas triaxial model will enable us to account for scatter in scaling relations between weak lensing mass and gas observables (e.g., tSZ-weak lensing mass relation) due to the triaxial shape of the DM halo and its correlation with the gas shape.

\section{Conclusions}\label{sec:summary}
In this work, we present X-ray Surface Brightness (XSB) and thermal Sunyaev-Zeldovich (tSZ) maps with the Baryon Pasting (BP) code, applying it to the half-sky lightcone derived from the Uchuu Cosmological $N$-body Simulations. These simulations encompass over $75$ million DM halos with masses $M_{500c} \geq 10^{13} M_\odot$, spanning a redshift range from $0$ to $2$. BP-Uchuu Maps facilitate the detection and evaluation of novel systematic effects in X-ray and SZ cosmological surveys at the map level. The vast sky coverage and large number of halos in the BP-Uchuu maps ensure that results are resilient to cosmic variance. 

We demonstrated that the triaxial shape of the halo gas significantly affects the scatter in the X-ray luminosity versus halo mass relationship. In particular, the triaxial gas contributes $8\%-9\%$ to the scatter in X-ray luminosity at a given mass for group and cluster-size halos with $M_{500c} \geq 5\times 10^{13} M_\odot$, constituting nearly half of the total scatter in core-excised X-ray luminosity.  This underscores the importance of its inclusion in standard cosmological analyses. 

We further showed that the intrinsic scatter in the thermal pressure and gas density profiles enhances the clustering power in both the X-ray and tSZ auto- and cross-angular power spectra. The scatter in halo profiles results in a $20\%$ increase in the X-ray auto-power spectrum at small multipoles $\ell = 100 $, and increases to $40\%$ at $\ell = 10^4$. The intrinsic scatter increases the tSZ auto-power spectrum by $25\%$ at $\ell = 100$, by $10\%$ at $\ell = 3000$, and by $20\%$ at $\ell = 10^4$. The X-ray/tSZ cross power spectrum is minimally impacted, with an increase in power by $25\%$ at $\ell = 100$, by $10\%$ at $\ell = 3000$, and by $15\%$ at $\ell = 10^4$. Ignoring this scatter in halo-model approaches could lead to biases in cosmological and astrophysical constraints with X-ray and tSZ power spectra.

The BP-Uchuu maps and halo catalog are available online for download via \href{https://app.globus.org/file-manager?origin_id=cf8dadb7-b6e9-4e2c-abc1-0813877efc13}{Globus}. The BP map-making code is also available upon request.

\section*{Acknowledgements}
We thank the anonymous referee for their comments and feedback. 
This work is supported by NASA ATP23-0154 grant and the Yale Center for Research Computing. 
A.F. acknowledges support from the National Science Foundation under Cooperative Agreement AST-2421782 the Simons Foundation award MPS-AI-00010515.
M.S. acknowledges support from MEXT KAKENHI Grant Number (20H05861, 23K19070, 24H00215, 24H00221).
KO is supported by JSPS KAKENHI Grant Number JP22K14036 and JP24H00215.
T.I. has been supported by IAAR Research Support Program in Chiba University Japan,
MEXT/JSPS KAKENHI (Grant Number JP19KK0344 and JP23H04002),
MEXT as ``Program for Promoting Researches on the Supercomputer Fugaku'' (JPMXP1020230406), and JICFuS.
HM is supported by JSPS KAKENHI Grand Numbers JP20H01932, JP23H00108, and 22K21349, and Tokai Pathways to Global Excellence (T-GEx), part of MEXT Strategic Professional Development Program for Young Researchers
We thank Instituto de Astrofisica de Andalucia (IAA-CSIC), Centro de Supercomputacion de Galicia (CESGA) and the Spanish academic and research network (RedIRIS) in Spain for hosting Uchuu DR1, DR2 and DR3 in the Skies \& Universes site for cosmological simulations. The Uchuu simulations were carried out on Aterui II supercomputer at Center for Computational Astrophysics, CfCA, of National Astronomical Observatory of Japan, and the K computer at the RIKEN Advanced Institute for Computational Science. The Uchuu Data Releases efforts have made use of the skun@IAA\_RedIRIS and skun6@IAA computer facilities managed by the IAA-CSIC in Spain (MICINN EU-Feder grant EQC2018-004366-P).

\software{
HEALpix \citep{healpix},
healpy \citep{healpy}
}

%%%%%%%%%%%%%%%%%%%%%%%%%%%%%%%%%%%%%%%%%%%%%%%%%%

%%%%%%%%%%%%%%%%%%%% REFERENCES %%%%%%%%%%%%%%%%%%

\bibliographystyle{aasjournal}
\bibliography{reference} 

\begin{thebibliography}{}
\expandafter\ifx\csname natexlab\endcsname\relax\def\natexlab#1{#1}\fi
\providecommand{\url}[1]{\href{#1}{#1}}
\providecommand{\dodoi}[1]{doi:~\href{http://doi.org/#1}{\nolinkurl{#1}}}
\providecommand{\doeprint}[1]{\href{http://ascl.net/#1}{\nolinkurl{http://ascl.net/#1}}}
\providecommand{\doarXiv}[1]{\href{https://arxiv.org/abs/#1}{\nolinkurl{https://arxiv.org/abs/#1}}}

\bibitem[{{Agarwal} {et~al.}(2018){Agarwal}, {Dav{\'e}}, \&
  {Bassett}}]{agarwal2018painting}
{Agarwal}, S., {Dav{\'e}}, R., \& {Bassett}, B.~A. 2018, \mnras, 478, 3410,
  \dodoi{10.1093/mnras/sty1169}

\bibitem[{{Allen} {et~al.}(2011){Allen}, {Evrard}, \& {Mantz}}]{allen_etal11}
{Allen}, S.~W., {Evrard}, A.~E., \& {Mantz}, A.~B. 2011, \araa, 49, 409,
  \dodoi{10.1146/annurev-astro-081710-102514}

\bibitem[{{Asplund} {et~al.}(2009){Asplund}, {Grevesse}, {Sauval}, \&
  {Scott}}]{asplund_etal09}
{Asplund}, M., {Grevesse}, N., {Sauval}, A.~J., \& {Scott}, P. 2009, \araa, 47,
  481, \dodoi{10.1146/annurev.astro.46.060407.145222}

\bibitem[{{Aung} {et~al.}(2021){Aung}, {Nagai}, \& {Lau}}]{aung_etal21}
{Aung}, H., {Nagai}, D., \& {Lau}, E.~T. 2021, \mnras, 508, 2071,
  \dodoi{10.1093/mnras/stab2598}

\bibitem[{{Aung} {et~al.}(2023){Aung}, {Nagai}, {Klypin}, {Behroozi},
  {Abdullah}, {Ishiyama}, {Prada}, {P{\'e}rez}, {L{\'o}pez Cacheiro}, \&
  {Ruedas}}]{Aung2023}
{Aung}, H., {Nagai}, D., {Klypin}, A., {et~al.} 2023, \mnras, 519, 1648,
  \dodoi{10.1093/mnras/stac3514}

\bibitem[{{Battaglia} {et~al.}(2012{\natexlab{a}}){Battaglia}, {Bond},
  {Pfrommer}, \& {Sievers}}]{battaglia_etal12}
{Battaglia}, N., {Bond}, J.~R., {Pfrommer}, C., \& {Sievers}, J.~L.
  2012{\natexlab{a}}, \apj, 758, 74, \dodoi{10.1088/0004-637X/758/2/74}

\bibitem[{{Battaglia} {et~al.}(2012{\natexlab{b}}){Battaglia}, {Bond},
  {Pfrommer}, \& {Sievers}}]{battaglia_etal12b}
---. 2012{\natexlab{b}}, \apj, 758, 75, \dodoi{10.1088/0004-637X/758/2/75}

\bibitem[{{Bayer} {et~al.}(2024){Bayer}, {Zhong}, {Li}, {DeRose}, {Feng}, \&
  {Liu}}]{bayer_etal24}
{Bayer}, A.~E., {Zhong}, Y., {Li}, Z., {et~al.} 2024, arXiv e-prints,
  arXiv:2407.17462, \dodoi{10.48550/arXiv.2407.17462}

\bibitem[{{Becker} \& {Kravtsov}(2011)}]{becker_kravtsov11}
{Becker}, M.~R., \& {Kravtsov}, A.~V. 2011, \apj, 740, 25,
  \dodoi{10.1088/0004-637X/740/1/25}

\bibitem[{Behroozi {et~al.}(2019)Behroozi, Wechsler, Hearin, \&
  Conroy}]{behroozi2019universemachine}
Behroozi, P., Wechsler, R.~H., Hearin, A.~P., \& Conroy, C. 2019, Monthly
  Notices of the Royal Astronomical Society, 488, 3143

\bibitem[{{Behroozi} {et~al.}(2013){Behroozi}, {Wechsler}, \&
  {Wu}}]{Behroozi2013}
{Behroozi}, P.~S., {Wechsler}, R.~H., \& {Wu}, H.-Y. 2013, \apj, 762, 109,
  \dodoi{10.1088/0004-637X/762/2/109}

\bibitem[{{Benson} {et~al.}(2013){Benson}, {de Haan}, {Dudley}, {Reichardt},
  {Aird}, {Andersson}, {Armstrong}, {Ashby}, {Bautz}, {Bayliss}, {Bazin},
  {Bleem}, {Brodwin}, {Carlstrom}, {Chang}, {Cho}, {Clocchiatti}, {Crawford},
  {Crites}, {Desai}, {Dobbs}, {Foley}, {Forman}, {George}, {Gladders},
  {Gonzalez}, {Halverson}, {Harrington}, {High}, {Holder}, {Holzapfel},
  {Hoover}, {Hrubes}, {Jones}, {Joy}, {Keisler}, {Knox}, {Lee}, {Leitch},
  {Liu}, {Lueker}, {Luong-Van}, {Mantz}, {Marrone}, {McDonald}, {McMahon},
  {Mehl}, {Meyer}, {Mocanu}, {Mohr}, {Montroy}, {Murray}, {Natoli}, {Padin},
  {Plagge}, {Pryke}, {Rest}, {Ruel}, {Ruhl}, {Saliwanchik}, {Saro}, {Sayre},
  {Schaffer}, {Shaw}, {Shirokoff}, {Song}, {Spieler}, {Stalder},
  {Staniszewski}, {Stark}, {Story}, {Stubbs}, {Suhada}, {van Engelen},
  {Vanderlinde}, {Vieira}, {Vikhlinin}, {Williamson}, {Zahn}, \&
  {Zenteno}}]{benson_etal13}
{Benson}, B.~A., {de Haan}, T., {Dudley}, J.~P., {et~al.} 2013, \apj, 763, 147,
  \dodoi{10.1088/0004-637X/763/2/147}

\bibitem[{{Bernyk} {et~al.}(2016){Bernyk}, {Croton}, {Tonini}, {Hodkinson},
  {Hassan}, {Garel}, {Duffy}, {Poole}, \& {Hegarty}}]{Bernyk2016}
{Bernyk}, M., {Croton}, D.~J., {Tonini}, C., {et~al.} 2016, \apjs, 223, 9,
  \dodoi{10.3847/0067-0049/223/1/9}

\bibitem[{{Blaizot} {et~al.}(2005){Blaizot}, {Wadadekar}, {Guiderdoni},
  {Colombi}, {Bertin}, {Bouchet}, {Devriendt}, \& {Hatton}}]{Blaizot2005}
{Blaizot}, J., {Wadadekar}, Y., {Guiderdoni}, B., {et~al.} 2005, \mnras, 360,
  159, \dodoi{10.1111/j.1365-2966.2005.09019.x}

\bibitem[{{Bocquet} {et~al.}(2019){Bocquet}, {Dietrich}, {Schrabback}, {Bleem},
  {Klein}, {Allen}, {Applegate}, {Ashby}, {Bautz}, {Bayliss}, {Benson},
  {Brodwin}, {Bulbul}, {Canning}, {Capasso}, {Carlstrom}, {Chang}, {Chiu},
  {Cho}, {Clocchiatti}, {Crawford}, {Crites}, {de Haan}, {Desai}, {Dobbs},
  {Foley}, {Forman}, {Garmire}, {George}, {Gladders}, {Gonzalez}, {Grandis},
  {Gupta}, {Halverson}, {Hlavacek-Larrondo}, {Hoekstra}, {Holder}, {Holzapfel},
  {Hou}, {Hrubes}, {Huang}, {Jones}, {Khullar}, {Knox}, {Kraft}, {Lee}, {von
  der Linden}, {Luong-Van}, {Mantz}, {Marrone}, {McDonald}, {McMahon}, {Meyer},
  {Mocanu}, {Mohr}, {Morris}, {Padin}, {Patil}, {Pryke}, {Rapetti},
  {Reichardt}, {Rest}, {Ruhl}, {Saliwanchik}, {Saro}, {Sayre}, {Schaffer},
  {Shirokoff}, {Stalder}, {Stanford}, {Staniszewski}, {Stark}, {Story},
  {Strazzullo}, {Stubbs}, {Vanderlinde}, {Vieira}, {Vikhlinin}, {Williamson},
  \& {Zenteno}}]{bocquet2019cluster}
{Bocquet}, S., {Dietrich}, J.~P., {Schrabback}, T., {et~al.} 2019, \apj, 878,
  55, \dodoi{10.3847/1538-4357/ab1f10}

\bibitem[{{Bode} {et~al.}(2009){Bode}, {Ostriker}, \&
  {Vikhlinin}}]{bode_etal09}
{Bode}, P., {Ostriker}, J.~P., \& {Vikhlinin}, A. 2009, \apj, 700, 989,
  \dodoi{10.1088/0004-637X/700/2/989}

\bibitem[{{Bolliet} {et~al.}(2018){Bolliet}, {Comis}, {Komatsu}, \&
  {Mac{\'\i}as-P{\'e}rez}}]{bolliet_etal18}
{Bolliet}, B., {Comis}, B., {Komatsu}, E., \& {Mac{\'\i}as-P{\'e}rez}, J.~F.
  2018, \mnras, 477, 4957, \dodoi{10.1093/mnras/sty823}

\bibitem[{Bryan \& Norman(1998)}]{bryan_norman98}
Bryan, G.~L., \& Norman, M.~L. 1998, \apj, 495, 80, \dodoi{10.1086/305262}

\bibitem[{{Chadayammuri} {et~al.}(2023){Chadayammuri}, {Ntampaka}, {ZuHone},
  {Bogd{\'a}n}, \& {Kraft}}]{chadayammuri_etal23}
{Chadayammuri}, U., {Ntampaka}, M., {ZuHone}, J., {Bogd{\'a}n}, {\'A}., \&
  {Kraft}, R.~P. 2023, \mnras, 526, 2812, \dodoi{10.1093/mnras/stad2596}

\bibitem[{{Clerc} {et~al.}(2018){Clerc}, {Ramos-Ceja}, {Ridl}, {Lamer},
  {Brunner}, {Hofmann}, {Comparat}, {Pacaud}, {K{\"a}fer}, {Reiprich},
  {Merloni}, {Schmid}, {Brand}, {Wilms}, {Friedrich}, {Finoguenov}, {Dauser},
  \& {Kreykenbohm}}]{clerc2018synthetic}
{Clerc}, N., {Ramos-Ceja}, M.~E., {Ridl}, J., {et~al.} 2018, \aap, 617, A92,
  \dodoi{10.1051/0004-6361/201732119}

\bibitem[{{Comparat} {et~al.}(2020){Comparat}, {Eckert}, {Finoguenov},
  {Schmidt}, {Sanders}, {Nagai}, {Lau}, {Kaefer}, {Pacaud}, {Clerc},
  {Reiprich}, {Bulbul}, {Chitham}, {Chiang}, {Ghirardini}, {Gonzalez-Perez},
  {Gozaliasl}, {Fitzpatrick}, {Klypin}, {Merloni}, {Nandra}, {Liu}, {Prada},
  {Ramos-Ceja}, {Salvato}, {Seppi}, {Tempel}, \& {Yepes}}]{comparat_etal20}
{Comparat}, J., {Eckert}, D., {Finoguenov}, A., {et~al.} 2020, The Open Journal
  of Astrophysics, 3, 13, \dodoi{10.21105/astro.2008.08404}

\bibitem[{{Costanzi} {et~al.}(2019){Costanzi}, {Rozo}, {Simet}, {Zhang},
  {Evrard}, {Mantz}, {Rykoff}, {Jeltema}, {Gruen}, {Allen}, {McClintock},
  {Romer}, {von der Linden}, {Farahi}, {DeRose}, {Varga}, {Weller}, {Giles},
  {Hollowood}, {Bhargava}, {Bermeo-Hernandez}, {Chen}, {Abbott}, {Abdalla},
  {Avila}, {Bechtol}, {Brooks}, {Buckley-Geer}, {Burke}, {Rosell}, {Kind},
  {Carretero}, {Crocce}, {Cunha}, {da Costa}, {Davis}, {De Vicente}, {Diehl},
  {Dietrich}, {Doel}, {Eifler}, {Estrada}, {Flaugher}, {Fosalba}, {Frieman},
  {Garc{\'\i}a-Bellido}, {Gaztanaga}, {Gerdes}, {Giannantonio}, {Gruendl},
  {Gschwend}, {Gutierrez}, {Hartley}, {Honscheid}, {Hoyle}, {James}, {Krause},
  {Kuehn}, {Kuropatkin}, {Lima}, {Lin}, {Maia}, {March}, {Marshall}, {Martini},
  {Menanteau}, {Miller}, {Miquel}, {Mohr}, {Ogando}, {Plazas}, {Roodman},
  {Sanchez}, {Scarpine}, {Schindler}, {Schubnell}, {Serrano}, {Sevilla-Noarbe},
  {Sheldon}, {Smith}, {Soares-Santos}, {Sobreira}, {Suchyta}, {Swanson},
  {Tarle}, {Thomas}, \& {Wechsler}}]{costanzi2019methods}
{Costanzi}, M., {Rozo}, E., {Simet}, M., {et~al.} 2019, \mnras, 488, 4779,
  \dodoi{10.1093/mnras/stz1949}

\bibitem[{{Diemer} \& {Kravtsov}(2015)}]{diemer_kravtsov15}
{Diemer}, B., \& {Kravtsov}, A.~V. 2015, \apj, 799, 108,
  \dodoi{10.1088/0004-637X/799/1/108}

\bibitem[{{Dong-P{\'a}ez} {et~al.}(2024){Dong-P{\'a}ez}, {Smith}, {Szewciw},
  {Ereza}, {Abdullah}, {Hern{\'a}ndez-Aguayo}, {Trusov}, {Prada}, {Klypin},
  {Ishiyama}, {Berlind}, {Zarrouk}, {L{\'o}pez Cacheiro}, \&
  {Ruedas}}]{Paez2024}
{Dong-P{\'a}ez}, C.~A., {Smith}, A., {Szewciw}, A.~O., {et~al.} 2024, \mnras,
  528, 7236, \dodoi{10.1093/mnras/stae062}

\bibitem[{{Dunkley} {et~al.}(2013){Dunkley}, {Calabrese}, {Sievers}, {Addison},
  {Battaglia}, {Battistelli}, {Bond}, {Das}, {Devlin}, {D{\"u}nner}, {Fowler},
  {Gralla}, {Hajian}, {Halpern}, {Hasselfield}, {Hincks}, {Hlozek}, {Hughes},
  {Irwin}, {Kosowsky}, {Louis}, {Marriage}, {Marsden}, {Menanteau}, {Moodley},
  {Niemack}, {Nolta}, {Page}, {Partridge}, {Sehgal}, {Spergel}, {Staggs},
  {Switzer}, {Trac}, \& {Wollack}}]{act_szpower}
{Dunkley}, J., {Calabrese}, E., {Sievers}, J., {et~al.} 2013, \jcap, 2013, 025,
  \dodoi{10.1088/1475-7516/2013/07/025}

\bibitem[{{Ereza} {et~al.}(2024){Ereza}, {Prada}, {Klypin}, {Ishiyama},
  {Smith}, {Baugh}, {Li}, {H\ ern{\'a}ndez-Aguayo}, \& {Ruedas}}]{Ereza2024}
{Ereza}, J., {Prada}, F., {Klypin}, A., {et~al.} 2024, \mnras, 532, 1659,
  \dodoi{10.1093/mnras/stae1543}

\bibitem[{Farahi {et~al.}(2022)Farahi, Anbajagane, \& Evrard}]{kllr}
Farahi, A., Anbajagane, D., \& Evrard, A.~E. 2022, The Astrophysical Journal,
  931, 166

\bibitem[{{Farahi} {et~al.}(2018){Farahi}, {Evrard}, {McCarthy}, {Barnes}, \&
  {Kay}}]{farahi_etal18}
{Farahi}, A., {Evrard}, A.~E., {McCarthy}, I., {Barnes}, D.~J., \& {Kay}, S.~T.
  2018, \mnras, 478, 2618, \dodoi{10.1093/mnras/sty1179}

\bibitem[{{Farahi} {et~al.}(2019{\natexlab{a}}){Farahi}, {Mulroy}, {Evrard},
  {Smith}, {Finoguenov}, {Bourdin}, {Carlstrom}, {Haines}, {Marrone},
  {Martino}, {Mazzotta}, {O'Donnell}, \& {Okabe}}]{farahi2019NatureCorrelation}
{Farahi}, A., {Mulroy}, S.~L., {Evrard}, A.~E., {et~al.} 2019{\natexlab{a}},
  Nature Communications, 10, 2504, \dodoi{10.1038/s41467-019-10471-y}

\bibitem[{{Farahi} {et~al.}(2019{\natexlab{b}}){Farahi}, {Chen}, {Evrard},
  {Hollowood}, {Wilkinson}, {Bhargava}, {Giles}, {Romer}, {Jeltema}, {Hilton},
  {Bermeo}, {Mayers}, {Vergara Cervantes}, {Rozo}, {Rykoff}, {Collins},
  {Costanzi}, {Everett}, {Liddle}, {Mann}, {Mantz}, {Rooney}, {Sahlen},
  {Stott}, {Viana}, {Zhang}, {Annis}, {Avila}, {Brooks}, {Buckley-Geer},
  {Burke}, {Carnero Rosell}, {Carrasco Kind}, {Carretero}, {Castander}, {da
  Costa}, {De Vicente}, {Desai}, {Diehl}, {Dietrich}, {Doel}, {Flaugher},
  {Fosalba}, {Frieman}, {Garc{\'\i}a-Bellido}, {Gaztanaga}, {Gerdes}, {Gruen},
  {Gruendl}, {Gschwend}, {Gutierrez}, {Honscheid}, {James}, {Krause}, {Kuehn},
  {Kuropatkin}, {Lima}, {Maia}, {Marshall}, {Melchior}, {Menanteau}, {Miquel},
  {Ogando}, {Plazas}, {Sanchez}, {Scarpine}, {Schubnell}, {Serrano},
  {Sevilla-Noarbe}, {Smith}, {Sobreira}, {Suchyta}, {Swanson}, {Tarle},
  {Thomas}, {Tucker}, {Vikram}, {Walker}, {Weller}, \& {DES
  Collaboration}}]{farahi2019mass}
{Farahi}, A., {Chen}, X., {Evrard}, A.~E., {et~al.} 2019{\natexlab{b}}, \mnras,
  490, 3341, \dodoi{10.1093/mnras/stz2689}

\bibitem[{{Fern{\'a}ndez-Garc{\'\i}a}
  {et~al.}(2024){Fern{\'a}ndez-Garc{\'\i}a}, {Betancort-Rijo}, {Prada},
  {Ishiyama}, \& {Klypin}}]{Garcia2024}
{Fern{\'a}ndez-Garc{\'\i}a}, E., {Betancort-Rijo}, J.~E., {Prada}, F.,
  {Ishiyama}, T., \& {Klypin}, A. 2024, arXiv e-prints, arXiv:2406.13736,
  \dodoi{10.48550/arXiv.2406.13736}

\bibitem[{{Flender} {et~al.}(2017){Flender}, {Nagai}, \&
  {McDonald}}]{flender_etal17}
{Flender}, S., {Nagai}, D., \& {McDonald}, M. 2017, \apj, 837, 124,
  \dodoi{10.3847/1538-4357/aa60bf}

\bibitem[{{Foster} {et~al.}(2012){Foster}, {Ji}, {Smith}, \&
  {Brickhouse}}]{foster_etal12}
{Foster}, A.~R., {Ji}, L., {Smith}, R.~K., \& {Brickhouse}, N.~S. 2012, \apj,
  756, 128, \dodoi{10.1088/0004-637X/756/2/128}

\bibitem[{{George} {et~al.}(2015){George}, {Reichardt}, {Aird}, {Benson},
  {Bleem}, {Carlstrom}, {Chang}, {Cho}, {Crawford}, {Crites}, {de Haan},
  {Dobbs}, {Dudley}, {Halverson}, {Harrington}, {Holder}, {Holzapfel}, {Hou},
  {Hrubes}, {Keisler}, {Knox}, {Lee}, {Leitch}, {Lueker}, {Luong-Van},
  {McMahon}, {Mehl}, {Meyer}, {Millea}, {Mocanu}, {Mohr}, {Montroy}, {Padin},
  {Plagge}, {Pryke}, {Ruhl}, {Schaffer}, {Shaw}, {Shirokoff}, {Spieler},
  {Staniszewski}, {Stark}, {Story}, {van Engelen}, {Vanderlinde}, {Vieira},
  {Williamson}, \& {Zahn}}]{spt_szpower}
{George}, E.~M., {Reichardt}, C.~L., {Aird}, K.~A., {et~al.} 2015, \apj, 799,
  177, \dodoi{10.1088/0004-637X/799/2/177}

\bibitem[{{Ghirardini} {et~al.}(2019){Ghirardini}, {Eckert}, {Ettori},
  {Pointecouteau}, {Molendi}, {Gaspari}, {Rossetti}, {De Grandi}, {Roncarelli},
  {Bourdin}, {Mazzotta}, {Rasia}, \& {Vazza}}]{ghirardini_etal19}
{Ghirardini}, V., {Eckert}, D., {Ettori}, S., {et~al.} 2019, \aap, 621, A41,
  \dodoi{10.1051/0004-6361/201833325}

\bibitem[{{Gkogkou} {et~al.}(2023){Gkogkou}, {B{\'e}thermin}, {Lagache}, {Van
  Cuyck}, {Jullo}, {Aravena}, {Beelen}, {Benoit}, {\ Bounmy}, {Calvo},
  {Catalano}, {Cora}, {Croton}, {de la Torre}, {Fasano}, {Ferrara}, {Goupy},
  {Hoarau}, {Hu}, {Ishiyama}, {Knudsen}, {Lambert}, {Mac{\'\i}as-P{\'e}rez},
  {Marpaud}, {Mellema}, {Monfardini}, {Pallottini}, {Ponthieu}, {Prada},
  {Roehlly}, {Vallini}, \& {Walter}}]{Gkogkou2023}
{Gkogkou}, A., {B{\'e}thermin}, M., {Lagache}, G., {et~al.} 2023, \aap, 670,
  A16, \dodoi{10.1051/0004-6361/202245151}

\bibitem[{{G{\'o}rski} {et~al.}(2005){G{\'o}rski}, {Hivon}, {Banday},
  {Wandelt}, {Hansen}, {Reinecke}, \& {Bartelmann}}]{healpix}
{G{\'o}rski}, K.~M., {Hivon}, E., {Banday}, A.~J., {et~al.} 2005, \apj, 622,
  759, \dodoi{10.1086/427976}

\bibitem[{{Hadzhiyska} {et~al.}(2020){Hadzhiyska}, {Bose}, {Eisenstein},
  {Hernquist}, \& {Spergel}}]{hadzhiyska2020limitations}
{Hadzhiyska}, B., {Bose}, S., {Eisenstein}, D., {Hernquist}, L., \& {Spergel},
  D.~N. 2020, \mnras, 493, 5506, \dodoi{10.1093/mnras/staa623}

\bibitem[{{He} {et~al.}(2021){He}, {Mansfield}, {Rau}, {Trac}, \&
  {Battaglia}}]{he_etal21}
{He}, Y., {Mansfield}, P., {Rau}, M.~M., {Trac}, H., \& {Battaglia}, N. 2021,
  \apj, 908, 91, \dodoi{10.3847/1538-4357/abd0ff}

\bibitem[{{Hurier} {et~al.}(2015){Hurier}, {Douspis}, {Aghanim},
  {Pointecouteau}, {Diego}, \& {Macias-Perez}}]{hurier_etal15}
{Hurier}, G., {Douspis}, M., {Aghanim}, N., {et~al.} 2015, \aap, 576, A90,
  \dodoi{10.1051/0004-6361/201425555}

\bibitem[{{Ishiyama} {et~al.}(2021){Ishiyama}, {Prada}, {Klypin}, {Sinha},
  {Metcalf}, {Jullo}, {Altieri}, {Cora}, {Croton}, {de la Torre},
  {Mill{\'a}n-Calero}, {Oogi}, {Ruedas}, \& {Vega-Mart{\'\i}nez}}]{uchuu}
{Ishiyama}, T., {Prada}, F., {Klypin}, A.~A., {et~al.} 2021, \mnras, 506, 4210,
  \dodoi{10.1093/mnras/stab1755}

\bibitem[{{Kazantzidis} {et~al.}(2004){Kazantzidis}, {Kravtsov}, {Zentner},
  {Allgood}, {Nagai}, \& {Moore}}]{kazantzidis_etal04}
{Kazantzidis}, S., {Kravtsov}, A.~V., {Zentner}, A.~R., {et~al.} 2004, \apjl,
  611, L73, \dodoi{10.1086/423992}

\bibitem[{{K{\'e}ruzor{\'e}} {et~al.}(2024){K{\'e}ruzor{\'e}}, {Bleem},
  {Frontiere}, {Krishnan}, {Buehlmann}, {Emberson}, {Habib}, \&
  {Larsen}}]{picasso}
{K{\'e}ruzor{\'e}}, F., {Bleem}, L.~E., {Frontiere}, N., {et~al.} 2024, The
  Open Journal of Astrophysics, 7, 116, \dodoi{10.33232/001c.127486}

\bibitem[{{Kim} {et~al.}(2024){Kim}, {Sayers}, {Sereno}, {Bartalucci},
  {Chappuis}, {De Grandi}, {De Luca}, {De Petris}, {Donahue}, {Eckert},
  {Ettori}, {Gaspari}, {Gastaldello}, {Gavazzi}, {Gavidia}, {Ghizzardi},
  {Iqbal}, {Kay}, {Lovisari}, {Maughan}, {Mazzotta}, {Okabe}, {Pointecouteau},
  {Pratt}, {Rossetti}, \& {Umetsu}}]{kim_etal24}
{Kim}, J., {Sayers}, J., {Sereno}, M., {et~al.} 2024, \aap, 686, A97,
  \dodoi{10.1051/0004-6361/202347399}

\bibitem[{{Komatsu} \& {Seljak}(2001)}]{komatsu_seljak01}
{Komatsu}, E., \& {Seljak}, U. 2001, \mnras, 327, 1353,
  \dodoi{10.1046/j.1365-8711.2001.04838.x}

\bibitem[{{Lau} {et~al.}(2023){Lau}, {Bogd{\'a}n}, {Chadayammuri}, {Nagai},
  {Kraft}, \& {Cappelluti}}]{lau_etal23}
{Lau}, E.~T., {Bogd{\'a}n}, {\'A}., {Chadayammuri}, U., {et~al.} 2023, \mnras,
  518, 1496, \dodoi{10.1093/mnras/stac3147}

\bibitem[{{Lau} {et~al.}(2024){Lau}, {Bogd\'an}, {Nagai}, {Cappelluti}, \&
  {Shirasaki}}]{lau_etal24}
{Lau}, E.~T., {Bogd\'an}, {\'A}., {Nagai}, D., {Cappelluti}, N., \&
  {Shirasaki}, M. 2024.
\newblock \doarXiv{2410.22397}

\bibitem[{{Lau} {et~al.}(2021){Lau}, {Hearin}, {Nagai}, \&
  {Cappelluti}}]{lau_etal2020}
{Lau}, E.~T., {Hearin}, A.~P., {Nagai}, D., \& {Cappelluti}, N. 2021, \mnras,
  500, 1029, \dodoi{10.1093/mnras/staa3313}

\bibitem[{{Lau} {et~al.}(2011){Lau}, {Nagai}, {Kravtsov}, \&
  {Zentner}}]{lau_etal11}
{Lau}, E.~T., {Nagai}, D., {Kravtsov}, A.~V., \& {Zentner}, A.~R. 2011, \apj,
  734, 93, \dodoi{10.1088/0004-637X/734/2/93}

\bibitem[{{Lee} {et~al.}(2024){Lee}, {Genel}, {Wandelt}, {Zhang}, {Delgado},
  {Pandey}, {Lau}, {Carr}, {Cook}, {Nagai}, {Angles-Alcazar},
  {Villaescusa-Navarro}, \& {Bryan}}]{carpoolgp}
{Lee}, M.~E., {Genel}, S., {Wandelt}, B.~D., {et~al.} 2024, \apj, 968, 11,
  \dodoi{10.3847/1538-4357/ad3d4a}

\bibitem[{{Limousin} {et~al.}(2013){Limousin}, {Morandi}, {Sereno},
  {Meneghetti}, {Ettori}, {Bartelmann}, \& {Verdugo}}]{limousin2013three}
{Limousin}, M., {Morandi}, A., {Sereno}, M., {et~al.} 2013, \ssr, 177, 155,
  \dodoi{10.1007/s11214-013-9980-y}

\bibitem[{{Machado} {et~al.}(2021){Machado}, {Avestruz}, {Barnes}, {Farahi},
  {Lau}, \& {Nagai}}]{machado_etal21}
{Machado}, L.~F., {Avestruz}, C., {Barnes}, D.~J., {et~al.} 2021, \mnras, 507,
  1468, \dodoi{10.1093/mnras/stab2252}

\bibitem[{Mantz {et~al.}(2010)Mantz, Allen, Ebeling, Rapetti, \&
  Drlica-Wagner}]{mantz2010observed}
Mantz, A., Allen, S.~W., Ebeling, H., Rapetti, D., \& Drlica-Wagner, A. 2010,
  Monthly Notices of the Royal Astronomical Society, 406, 1773

\bibitem[{{Mantz} {et~al.}(2016){Mantz}, {Allen}, {Morris}, {von der Linden},
  {Applegate}, {Kelly}, {Burke}, {Donovan}, \& {Ebeling}}]{mantz2016weighing}
{Mantz}, A.~B., {Allen}, S.~W., {Morris}, R.~G., {et~al.} 2016, \mnras, 463,
  3582, \dodoi{10.1093/mnras/stw2250}

\bibitem[{Mead {et~al.}(2021)Mead, Brieden, Tr{\"o}ster, \&
  Heymans}]{mead2021hmcode}
Mead, A., Brieden, S., Tr{\"o}ster, T., \& Heymans, C. 2021, Monthly Notices of
  the Royal Astronomical Society, 502, 1401

\bibitem[{{Medlock} {et~al.}(2024){Medlock}, {Neufeld}, {Nagai}, {Angl{\'e}s
  Alc{\'a}zar}, {Genel}, {Oppenheimer}, {Singh}, \&
  {Villaescusa-Navarro}}]{medlock_etal24}
{Medlock}, I., {Neufeld}, C., {Nagai}, D., {et~al.} 2024, arXiv e-prints,
  arXiv:2410.16361, \dodoi{10.48550/arXiv.2410.16361}

\bibitem[{{Melin} \& {Pratt}(2023)}]{melin_pratt23}
{Melin}, J.~B., \& {Pratt}, G.~W. 2023, \aap, 678, A197,
  \dodoi{10.1051/0004-6361/202346690}

\bibitem[{{Mernier} {et~al.}(2018){Mernier}, {Biffi}, {Yamaguchi}, {Medvedev},
  {Simionescu}, {Ettori}, {Werner}, {Kaastra}, {de Plaa}, \&
  {Gu}}]{mernier_etal18}
{Mernier}, F., {Biffi}, V., {Yamaguchi}, H., {et~al.} 2018, \ssr, 214, 129,
  \dodoi{10.1007/s11214-018-0565-7}

\bibitem[{{More} {et~al.}(2015){More}, {Diemer}, \& {Kravtsov}}]{more_etal15}
{More}, S., {Diemer}, B., \& {Kravtsov}, A.~V. 2015, \apj, 810, 36,
  \dodoi{10.1088/0004-637X/810/1/36}

\bibitem[{Moster {et~al.}(2018)Moster, Naab, \& White}]{moster2018emerge}
Moster, B.~P., Naab, T., \& White, S.~D. 2018, Monthly Notices of the Royal
  Astronomical Society, 477, 1822

\bibitem[{Mulroy {et~al.}(2019)Mulroy, Farahi, Evrard, Smith, Finoguenov,
  O’Donnell, Marrone, Abdulla, Bourdin, Carlstrom,
  {et~al.}}]{mulroy2019locuss}
Mulroy, S.~L., Farahi, A., Evrard, A.~E., {et~al.} 2019, Monthly Notices of the
  Royal Astronomical Society, 484, 60

\bibitem[{{Navarro} {et~al.}(1996){Navarro}, {Frenk}, \& {White}}]{nfw96}
{Navarro}, J.~F., {Frenk}, C.~S., \& {White}, S. D.~M. 1996, \apj, 462, 563,
  \dodoi{10.1086/177173}

\bibitem[{{Nelson} {et~al.}(2019){Nelson}, {Springel}, {Pillepich},
  {Rodriguez-Gomez}, {Torrey}, {Genel}, {Vogelsberger}, {Pakmor}, {Marinacci},
  {Weinberger}, {Kelley}, {Lovell}, {Diemer}, \& {Hernquist}}]{TNG}
{Nelson}, D., {Springel}, V., {Pillepich}, A., {et~al.} 2019, Computational
  Astrophysics and Cosmology, 6, 2, \dodoi{10.1186/s40668-019-0028-x}

\bibitem[{{Nelson} {et~al.}(2014{\natexlab{a}}){Nelson}, {Lau}, \&
  {Nagai}}]{nelson_etal14b}
{Nelson}, K., {Lau}, E.~T., \& {Nagai}, D. 2014{\natexlab{a}}, \apj, 792, 25,
  \dodoi{10.1088/0004-637X/792/1/25}

\bibitem[{{Nelson} {et~al.}(2014{\natexlab{b}}){Nelson}, {Lau}, {Nagai},
  {Rudd}, \& {Yu}}]{nelson_etal14}
{Nelson}, K., {Lau}, E.~T., {Nagai}, D., {Rudd}, D.~H., \& {Yu}, L.
  2014{\natexlab{b}}, \apj, 782, 107, \dodoi{10.1088/0004-637X/782/2/107}

\bibitem[{{Nguyen} {et~al.}(2024){Nguyen}, {Villaescusa-Navarro},
  {Mishra-Sharma}, {Cuesta-Lazaro}, {Torrey}, {Farahi}, {Garcia}, {Rose},
  {O'Neil}, {Vogelsberger}, {Shen}, {Roche}, {Angl{\'e}s-Alc{\'a}zar},
  {Kallivayalil}, {Mu{\~n}oz}, {Cyr-Racine}, {Roy}, {Necib}, \&
  {Kollmann}}]{nguyen2024dreams}
{Nguyen}, T., {Villaescusa-Navarro}, F., {Mishra-Sharma}, S., {et~al.} 2024,
  arXiv e-prints, arXiv:2409.02980, \dodoi{10.48550/arXiv.2409.02980}

\bibitem[{{Omori}(2022)}]{agora}
{Omori}, Y. 2022, arXiv e-prints, arXiv:2212.07420,
  \dodoi{10.48550/arXiv.2212.07420}

\bibitem[{{Oogi} {et~al.}(2023){Oogi}, {Ishiyama}, {Prada}, {Sinha}, {Croton},
  {Cora}, {Jullo}, {Nagashima}, {L{\'o}pez Cacheiro}, {Ruedas}, {Kobayashi}, \&
  {Makiya}}]{Oogi2023}
{Oogi}, T., {Ishiyama}, T., {Prada}, F., {et~al.} 2023, \mnras, 525, 3879,
  \dodoi{10.1093/mnras/stad2401}

\bibitem[{{Osato} {et~al.}(2018){Osato}, {Flender}, {Nagai}, {Shirasaki}, \&
  {Yoshida}}]{osato_etal18}
{Osato}, K., {Flender}, S., {Nagai}, D., {Shirasaki}, M., \& {Yoshida}, N.
  2018, \mnras, 475, 532, \dodoi{10.1093/mnras/stx3215}

\bibitem[{{Osato} \& {Nagai}(2023)}]{bp_algo}
{Osato}, K., \& {Nagai}, D. 2023, \mnras, 519, 2069,
  \dodoi{10.1093/mnras/stac3669}

\bibitem[{{Ostriker} {et~al.}(2005){Ostriker}, {Bode}, \&
  {Babul}}]{ostriker_etal05}
{Ostriker}, J.~P., {Bode}, P., \& {Babul}, A. 2005, \apj, 634, 964,
  \dodoi{10.1086/497122}

\bibitem[{{Planck Collaboration} {et~al.}(2013){Planck Collaboration}, {Ade},
  {Aghanim}, {Arnaud}, {Ashdown}, {Atrio-Barandela}, {Aumont}, {Baccigalupi},
  {Balbi}, {Banday}, \& et~al.}]{planck_pressure}
{Planck Collaboration}, {Ade}, P.~A.~R., {Aghanim}, N., {et~al.} 2013, \aap,
  550, A131, \dodoi{10.1051/0004-6361/201220040}

\bibitem[{{Planck Collaboration} {et~al.}(2016){Planck Collaboration},
  {Aghanim}, {Arnaud}, {Ashdown}, {Aumont}, {Baccigalupi}, {Banday},
  {Barreiro}, {Bartlett}, {Bartolo}, \& et~al.}]{planck_tsz}
{Planck Collaboration}, {Aghanim}, N., {Arnaud}, M., {et~al.} 2016, \aap, 594,
  A22, \dodoi{10.1051/0004-6361/201525826}

\bibitem[{{Planck Collaboration} {et~al.}(2020){Planck Collaboration},
  {Aghanim}, {Akrami}, {Ashdown}, {Aumont}, {Baccigalupi}, {Ballardini},
  {Banday}, {Barreiro}, {Bartolo}, {Basak}, {Battye}, {Benabed}, {Bernard},
  {Bersanelli}, {Bielewicz}, {Bock}, {Bond}, {Borrill}, {Bouchet}, {Boulanger},
  {Bucher}, {Burigana}, {Butler}, {Calabrese}, {Cardoso}, {Carron},
  {Challinor}, {Chiang}, {Chluba}, {Colombo}, {Combet}, {Contreras}, {Crill},
  {Cuttaia}, {de Bernardis}, {de Zotti}, {Delabrouille}, {Delouis}, {Di
  Valentino}, {Diego}, {Dor{\'e}}, {Douspis}, {Ducout}, {Dupac}, {Dusini},
  {Efstathiou}, {Elsner}, {En{\ss}lin}, {Eriksen}, {Fantaye}, {Farhang},
  {Fergusson}, {Fernandez-Cobos}, {Finelli}, {Forastieri}, {Frailis},
  {Fraisse}, {Franceschi}, {Frolov}, {Galeotta}, {Galli}, {Ganga},
  {G{\'e}nova-Santos}, {Gerbino}, {Ghosh}, {Gonz{\'a}lez-Nuevo}, {G{\'o}rski},
  {Gratton}, {Gruppuso}, {Gudmundsson}, {Hamann}, {Handley}, {Hansen},
  {Herranz}, {Hildebrandt}, {Hivon}, {Huang}, {Jaffe}, {Jones}, {Karakci},
  {Keih{\"a}nen}, {Keskitalo}, {Kiiveri}, {Kim}, {Kisner}, {Knox},
  {Krachmalnicoff}, {Kunz}, {Kurki-Suonio}, {Lagache}, {Lamarre}, {Lasenby},
  {Lattanzi}, {Lawrence}, {Le Jeune}, {Lemos}, {Lesgourgues}, {Levrier},
  {Lewis}, {Liguori}, {Lilje}, {Lilley}, {Lindholm}, {L{\'o}pez-Caniego},
  {Lubin}, {Ma}, {Mac{\'\i}as-P{\'e}rez}, {Maggio}, {Maino}, {Mandolesi},
  {Mangilli}, {Marcos-Caballero}, {Maris}, {Martin}, {Martinelli},
  {Mart{\'\i}nez-Gonz{\'a}lez}, {Matarrese}, {Mauri}, {McEwen}, {Meinhold},
  {Melchiorri}, {Mennella}, {Migliaccio}, {Millea}, {Mitra},
  {Miville-Desch{\^e}nes}, {Molinari}, {Montier}, {Morgante}, {Moss}, {Natoli},
  {N{\o}rgaard-Nielsen}, {Pagano}, {Paoletti}, {Partridge}, {Patanchon},
  {Peiris}, {Perrotta}, {Pettorino}, {Piacentini}, {Polastri}, {Polenta},
  {Puget}, {Rachen}, {Reinecke}, {Remazeilles}, {Renzi}, {Rocha}, {Rosset},
  {Roudier}, {Rubi{\~n}o-Mart{\'\i}n}, {Ruiz-Granados}, {Salvati}, {Sandri},
  {Savelainen}, {Scott}, {Shellard}, {Sirignano}, {Sirri}, {Spencer},
  {Sunyaev}, {Suur-Uski}, {Tauber}, {Tavagnacco}, {Tenti}, {Toffolatti},
  {Tomasi}, {Trombetti}, {Valenziano}, {Valiviita}, {Van Tent}, {Vibert},
  {Vielva}, {Villa}, {Vittorio}, {Wandelt}, {Wehus}, {White}, {White},
  {Zacchei}, \& {Zonca}}]{planck2018}
{Planck Collaboration}, {Aghanim}, N., {Akrami}, Y., {et~al.} 2020, \aap, 641,
  A6, \dodoi{10.1051/0004-6361/201833910}

\bibitem[{{Pop} {et~al.}(2022){Pop}, {Hernquist}, {Nagai}, {Kannan},
  {Weinberger}, {Springel}, {Vogelsberger}, {Nelson}, {Pakmor}, {Pillepich}, \&
  {Torrey}}]{pop_etal22}
{Pop}, A.-R., {Hernquist}, L., {Nagai}, D., {et~al.} 2022, arXiv e-prints,
  arXiv:2205.11528, \dodoi{10.48550/arXiv.2205.11528}

\bibitem[{{Prada} {et~al.}(2023){Prada}, {Behroozi}, {Ishiyama}, {Klypin}, \&
  {P{\'e}rez}}]{Prada2023}
{Prada}, F., {Behroozi}, P., {Ishiyama}, T., {Klypin}, A., \& {P{\'e}rez}, E.
  2023, arXiv e-prints, arXiv:2304.11911, \dodoi{10.48550/arXiv.2304.11911}

\bibitem[{{Pratt} {et~al.}(2019){Pratt}, {Arnaud}, {Biviano}, {Eckert},
  {Ettori}, {Nagai}, {Okabe}, \& {Reiprich}}]{pratt_etal19}
{Pratt}, G.~W., {Arnaud}, M., {Biviano}, A., {et~al.} 2019, \ssr, 215, 25,
  \dodoi{10.1007/s11214-019-0591-0}

\bibitem[{{Schaye} {et~al.}(2023){Schaye}, {Kugel}, {Schaller}, {Helly},
  {Braspenning}, {Elbers}, {McCarthy}, {van Daalen}, {Vandenbroucke}, {Frenk},
  {Kwan}, {Salcido}, {Bah{\'e}}, {Borrow}, {Chaikin}, {Hahn}, {Hu{\v{s}}ko},
  {Jenkins}, {Lacey}, \& {Nobels}}]{flamingo}
{Schaye}, J., {Kugel}, R., {Schaller}, M., {et~al.} 2023, \mnras, 526, 4978,
  \dodoi{10.1093/mnras/stad2419}

\bibitem[{{Seljak}(2000)}]{seljak2000analytic}
{Seljak}, U. 2000, \mnras, 318, 203, \dodoi{10.1046/j.1365-8711.2000.03715.x}

\bibitem[{{Sereno} {et~al.}(2020){Sereno}, {Umetsu}, {Ettori}, {Eckert},
  {Gastaldello}, {Giles}, {Lieu}, {Maughan}, {Okabe}, {Birkinshaw}, {Chiu},
  {Fujita}, {Miyazaki}, {Rapetti}, {Koulouridis}, \& {Pierre}}]{sereno2020xxl}
{Sereno}, M., {Umetsu}, K., {Ettori}, S., {et~al.} 2020, \mnras, 492, 4528,
  \dodoi{10.1093/mnras/stz3425}

\bibitem[{{Shaw} {et~al.}(2010){Shaw}, {Nagai}, {Bhattacharya}, \&
  {Lau}}]{shaw_etal10}
{Shaw}, L.~D., {Nagai}, D., {Bhattacharya}, S., \& {Lau}, E.~T. 2010, \apj,
  725, 1452, \dodoi{10.1088/0004-637X/725/2/1452}

\bibitem[{{Shirasaki} {et~al.}(2020){Shirasaki}, {Lau}, \&
  {Nagai}}]{shirasaki_etal20}
{Shirasaki}, M., {Lau}, E.~T., \& {Nagai}, D. 2020, \mnras, 491, 235,
  \dodoi{10.1093/mnras/stz3021}

\bibitem[{{Stark}(1977)}]{stark77}
{Stark}, A.~A. 1977, \apj, 213, 368, \dodoi{10.1086/155164}

\bibitem[{{Stein} {et~al.}(2020){Stein}, {Alvarez}, {Bond}, {van Engelen}, \&
  {Battaglia}}]{websky}
{Stein}, G., {Alvarez}, M.~A., {Bond}, J.~R., {van Engelen}, A., \&
  {Battaglia}, N. 2020, \jcap, 2020, 012, \dodoi{10.1088/1475-7516/2020/10/012}

\bibitem[{{Trac} {et~al.}(2011){Trac}, {Bode}, \& {Ostriker}}]{trac_etal11}
{Trac}, H., {Bode}, P., \& {Ostriker}, J.~P. 2011, \apj, 727, 94,
  \dodoi{10.1088/0004-637X/727/2/94}

\bibitem[{{Truong} {et~al.}(2018){Truong}, {Rasia}, {Mazzotta}, {Planelles},
  {Biffi}, {Fabjan}, {Beck}, {Borgani}, {Dolag}, {Gaspari}, {Granato},
  {Murante}, {Ragone-Figueroa}, \& {Steinborn}}]{Truong2018scatter_simulations}
{Truong}, N., {Rasia}, E., {Mazzotta}, P., {et~al.} 2018, \mnras, 474, 4089,
  \dodoi{10.1093/mnras/stx2927}

\bibitem[{{Valotti} {et~al.}(2018){Valotti}, {Pierre}, {Farahi}, {Evrard},
  {Faccioli}, {Sauvageot}, {Clerc}, \& {Pacaud}}]{valotti2018cosmological}
{Valotti}, A., {Pierre}, M., {Farahi}, A., {et~al.} 2018, \aap, 614, A72,
  \dodoi{10.1051/0004-6361/201731445}

\bibitem[{{Vikhlinin} {et~al.}(2009){Vikhlinin}, {Kravtsov}, {Burenin},
  {Ebeling}, {Forman}, {Hornstrup}, {Jones}, {Murray}, {Nagai}, {Quintana}, \&
  {Voevodkin}}]{vikhlinin_etal09}
{Vikhlinin}, A., {Kravtsov}, A.~V., {Burenin}, R.~A., {et~al.} 2009, \apj, 692,
  1060, \dodoi{10.1088/0004-637X/692/2/1060}

\bibitem[{{Villaescusa-Navarro} {et~al.}(2021){Villaescusa-Navarro},
  {Angl{\'e}s-Alc{\'a}zar}, {Genel}, {Spergel}, {Somerville}, {Dave},
  {Pillepich}, {Hernquist}, {Nelson}, {Torrey}, {Narayanan}, {Li}, {Philcox},
  {La Torre}, {Maria Delgado}, {Ho}, {Hassan}, {Burkhart}, {Wadekar},
  {Battaglia}, {Contardo}, \& {Bryan}}]{camels}
{Villaescusa-Navarro}, F., {Angl{\'e}s-Alc{\'a}zar}, D., {Genel}, S., {et~al.}
  2021, \apj, 915, 71, \dodoi{10.3847/1538-4357/abf7ba}

\bibitem[{{Wechsler} \& {Tinker}(2018)}]{wechsler_tinker18}
{Wechsler}, R.~H., \& {Tinker}, J.~L. 2018, \araa, 56, 435,
  \dodoi{10.1146/annurev-astro-081817-051756}

\bibitem[{{Williams} {et~al.}(2023){Williams}, {Khan}, \&
  {McQuinn}}]{williams_etal23}
{Williams}, I.~M., {Khan}, A., \& {McQuinn}, M. 2023, \mnras, 520, 3626,
  \dodoi{10.1093/mnras/stad293}

\bibitem[{{Yang} {et~al.}(2022){Yang}, {Cai}, {Cui}, {Dav{\'e}}, {Peacock}, \&
  {Sorini}}]{yang_etal22}
{Yang}, T., {Cai}, Y.-C., {Cui}, W., {et~al.} 2022, \mnras, 516, 4084,
  \dodoi{10.1093/mnras/stac2505}

\bibitem[{{Zandanel} {et~al.}(2018){Zandanel}, {Fornasa}, {Prada}, {Reiprich},
  {Pacaud}, \& {Klypin}}]{zandanel_etal18}
{Zandanel}, F., {Fornasa}, M., {Prada}, F., {et~al.} 2018, \mnras, 480, 987,
  \dodoi{10.1093/mnras/sty1901}

\bibitem[{{Zhang} {et~al.}(2024){Zhang}, {Farahi}, {Nagai}, {Lau}, {Frieman},
  {Ricci}, {von der Linden}, {Wu}, \& {LSST Dark Energy Science
  Collaboration}}]{zhang2024bias}
{Zhang}, Z., {Farahi}, A., {Nagai}, D., {et~al.} 2024, \mnras, 530, 3127,
  \dodoi{10.1093/mnras/stae999}

\bibitem[{{Zonca} {et~al.}(2019){Zonca}, {Singer}, {Lenz}, {Reinecke},
  {Rosset}, {Hivon}, \& {Gorski}}]{healpy}
{Zonca}, A., {Singer}, L., {Lenz}, D., {et~al.} 2019, The Journal of Open
  Source Software, 4, 1298, \dodoi{10.21105/joss.01298}

\end{thebibliography}

%%%%%%%%%%%%%%%%%%%%%%%%%%%%%%%%%%%%%%%%%%%%%%%%%%

%%%%%%%%%%%%%%%%% APPENDICES %%%%%%%%%%%%%%%%%%%%%
\begin{appendix}

\section{Validation of the BP model with Observations and Simulations}\label{sec:model_accuracy}

\subsection{Profile Comparison}

In this section, we compare the gas density and pressure profiles of the fiducial BP model against cosmological simulations and observations. 

\begin{figure}
    \centering
    \includegraphics[width=1\columnwidth]{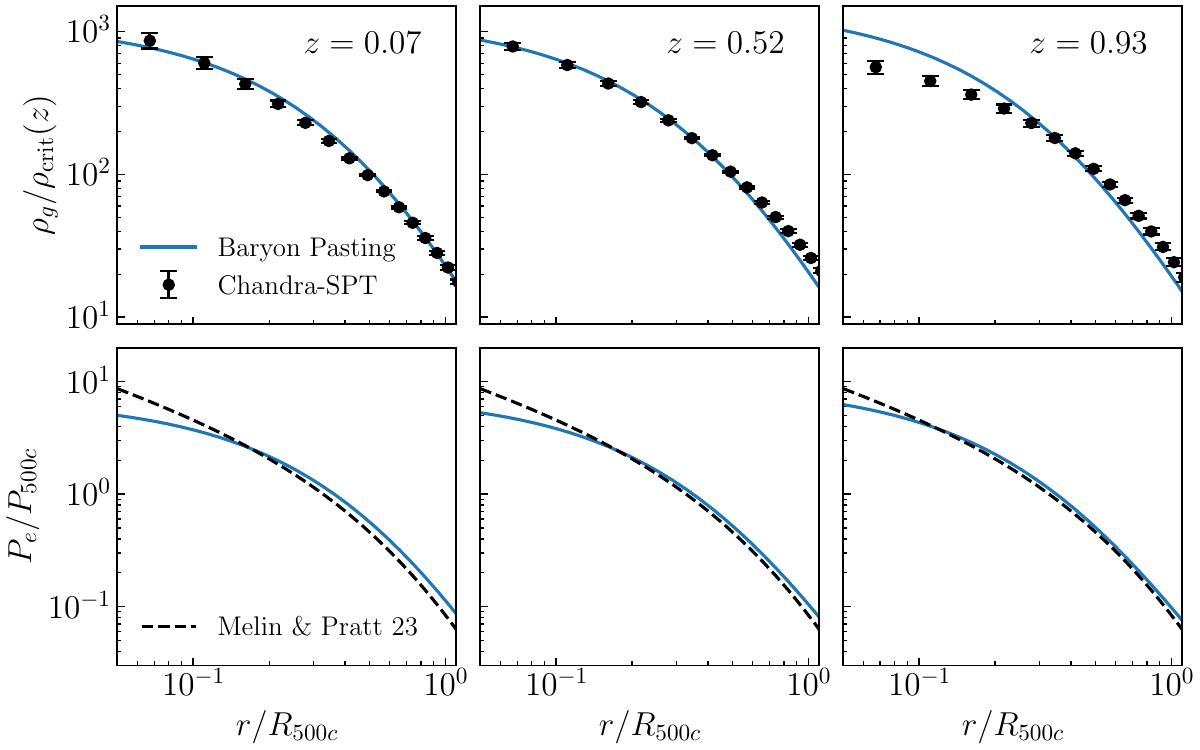}
    \caption{
    Comparison of profiles of gas density (top panels) and electron thermal pressure (bottom panels) with the BP model with default gas parameters for $M_{500c} = 5\times 10^{14} M_\odot$ at $z=0.07, 0.52, 0.93$ (left, middle, right panels), with the empiricl fit to to the Pressure profile from {\em Planck} and SPT-SZ data \citep{melin_pratt23} (dashed black line), and the density profile measurements from {\em Chandra}-SPT cluster sample presented in \citet{flender_etal17} as black data points with 1$\sigma$ errorbars.  
    }\label{fig:cluster_profiles_obs}
\end{figure}

\begin{figure}
    \centering
    \includegraphics[width=1\columnwidth]{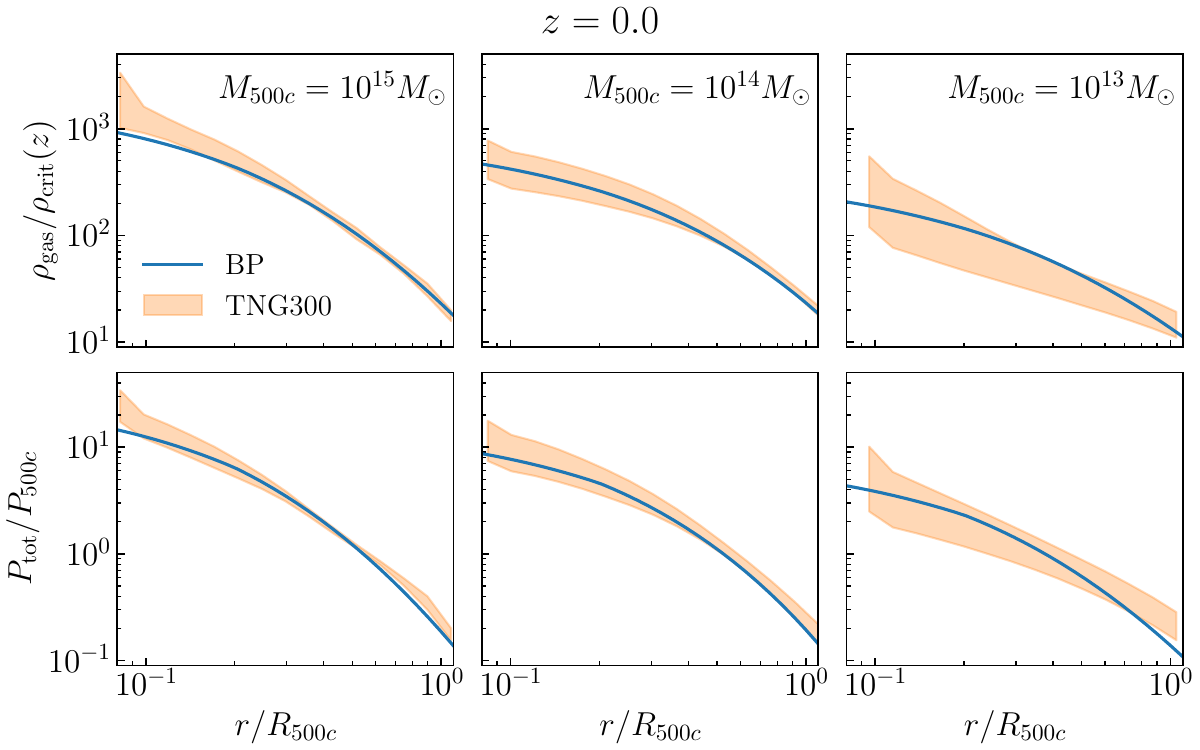}
    \caption{
    Comparision of the best-fit BP model (blue solid lines) to the gas density (top panels) and thermal pressure (bottom panels) of the TNG300 simulations for $\log _{10}(M_{500c}/M_\odot) =13,14,15$ at $z=0$.  The shaded regions represent $\pm 1\sigma$ of around the mean profile of the TNG300 halos. 
    }\label{fig:cluster_profiles_tng_mass}
\end{figure}

\begin{figure}
    \centering
    \includegraphics[width=1\columnwidth]{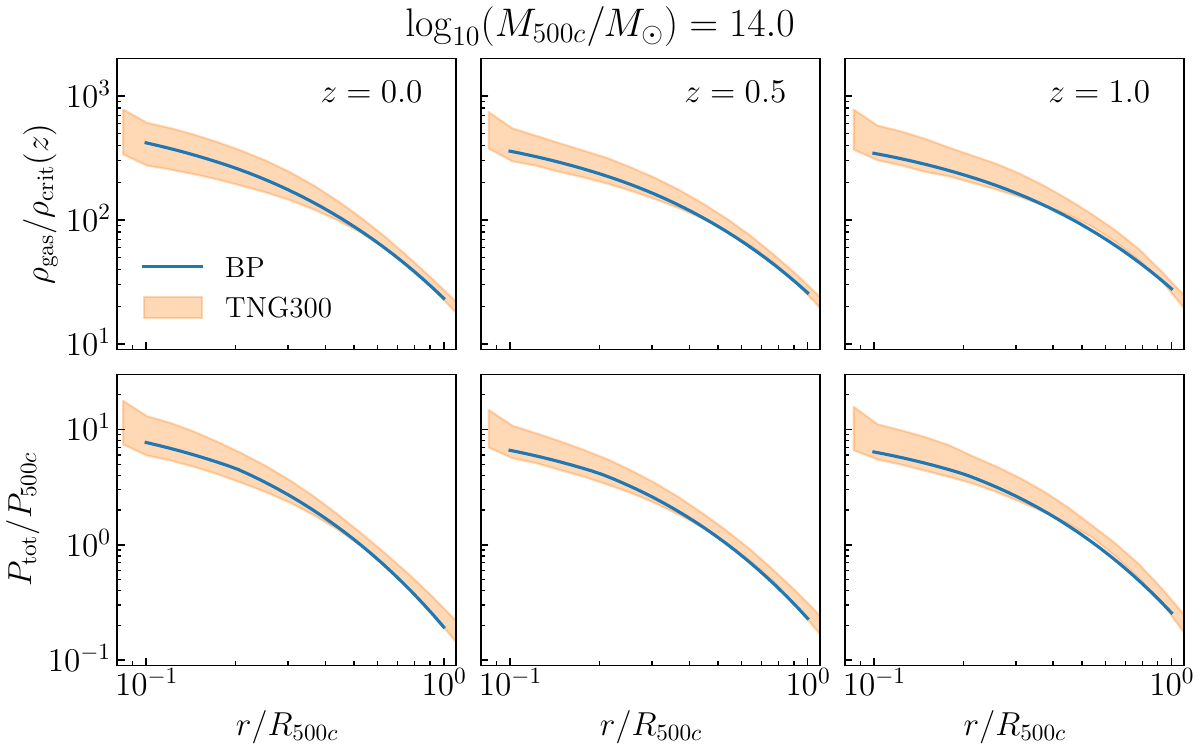}
    \caption{
    Same as Figure~\ref{fig:cluster_profiles_tng_mass}, but for $\log_{10}(M_{500c}/M_\odot)=14.0$ $z=0.0,0.5,1.0$ (from left to right). 
    }\label{fig:cluster_profiles_tng_z}
\end{figure}

Figure~\ref{fig:cluster_profiles_obs} shows the comparison of thermal pressure and gas density profiles of the BP Model to observations. In the top panels, we show the comparison between average gas density profile of clusters at $z=0.07, 0.52, 0.93$ from the {\em Chandra}-SPT cluster sample, to that of the BP model of a halo with mass $M_{500c} = 6\times 10^{14} M_\odot$. There is a very good agreement between the BP model and the {\em Chandra} measurements. This is not surprising given that the BP model is calibrated with the density profiles of the {\em Chandra}-SPT clusters \citep{flender_etal17}. 

In the bottom panels of the same figure, we compare the pressure profile between the fit to observations from {\em Planck} and SPT-SZ (\citealt{melin_pratt23}, see also \citealt{he_etal21}) for another updated model fit to the {\em Planck} measurements \citealt{planck_pressure}), and the BP model for the same halos. Again there is good agreement between the profiles of the BP model and the Universal Pressure Profile. Note that we do not fit the BP model to the {\em Planck}-XMM data. All the model parameters are from our calibration with the {\em Chandra}-SPT data, except the inner Polytropic index where we change from the fiducial value of $0.10$ to $1.0$. Note that the thermal pressure profiles are normalized by $P_{500c}  = 1.45 \times 10^{-11}\,{\rm erg\, cm^{-3}} \left(\frac{f_b}{0.174}\right) \left(\frac{h}{0.7}\right)^2 E(z)^{8/3} \left(\frac{M_{500c}}{10^{15} h^{-1}M_\odot}\right)^{2/3}$ to account for self-similar mass and redshift dependence in the normalization of the pressure profile. 

Figure~\ref{fig:cluster_profiles_tng_mass} shows the comparison of thermal pressure and gas density profiles from the TNG300 simulation \citep{TNG} at different halo masses $\log_{10}(M_{500c}/M_\odot)=(13, 14, 15)$ at $z=0$, and the corresponding profiles of the best-fit BP model. Figure~\ref{fig:cluster_profiles_tng_z} shows the same profiles for $\log_{10}(M_{500c}/M_\odot)=14$ at $z=0,0.5,1$.  Note that we fix the $f_\star = 0.019$ and $S_\star=0.16$ to those corresponding to the TNG300 simulations. The good agreement between the simulation and the BP model profiles for wide range of halo masses and redshifts demonstrates the flexibility of the BP model in describing thermodynamic profiles in cosmological simulations. 

\subsection{tSZ angular power spectrum}

\begin{figure}
\centering
    \includegraphics[width=1.0\columnwidth]{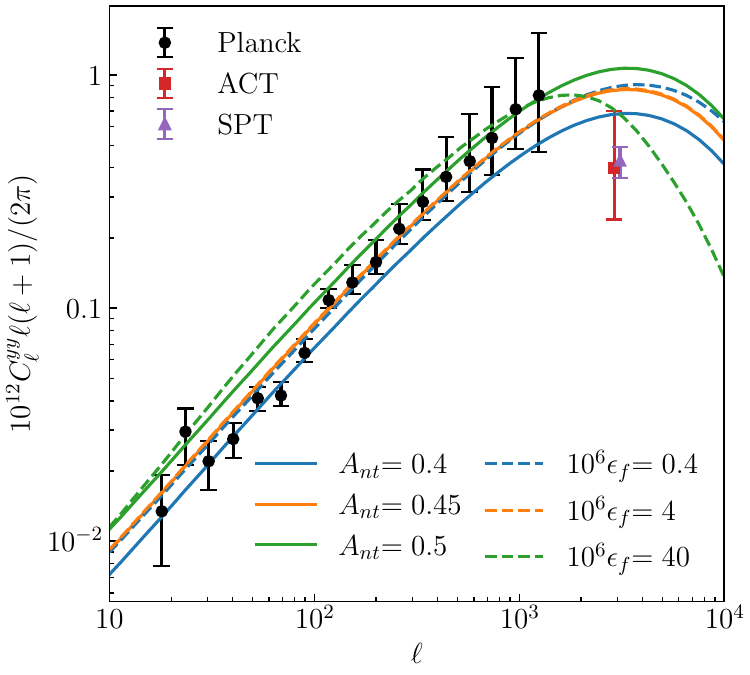}
    \caption{tSZ angular power spectra using the halo model in BP, with varying non-thermal pressure fraction parameter $A_{\rm nt}$. We also plot the {\em Planck} tSZ power spectrum \citep{planck_tsz} in black, ACT \citep{act_szpower} in red, and SPT \citep{spt_szpower} in purple, with 1$\sigma$ errorbars for comparison. The model power spectrum with the fiducial $A_{\rm nt} =0.45 $ and $10^6 \epsilon_f = 4.0$ agrees well with the mean power spectrum from {\em Planck}. The observation uncertainties can be bracketed with model power spectra with $A_{\rm nt} =\{0.40,0.50\} $. 
    }
    \label{fig:yy_power}
\end{figure} 

In Figure~\ref{fig:yy_power} we compare the Halo model-based tSZ power spectra computed with the BP code to the tSZ power spectrum from {\em Planck} \citep{planck_tsz}, ACT, and SPT. The BP tSZ power spectrum is computed with mass range $\log_{10}(M_{\rm vir}/M_\odot)=[13,16]$ and redshift range $z=[0.0,2.0]$, following Equations~(39)-(42) in \citet{bp_algo}, with varying non-thermal pressure fraction parameter $A_{\rm nt}$. We use the Planck 2018 cosmology \citep{planck2018} to compute the power spectrum. We compare the model power spectra with that of {\em Planck} \citep{planck_tsz}, ACT \citep{act_szpower}, and SPT \citep{spt_szpower}. The BP tSZ power spectrum with the fiducial non-thermal pressure fraction parameter $A_{\rm nt} = 0.45$ calibrated from cosmological hydrodynamical simulation provides a good match to the {\em Planck} measurements. This is also consistent with previous works that compared the BP model power spectra with observations \citep{osato_etal18}.

\end{appendix}
%%%%%%%%%%%%%%%%%%%%%%%%%%%%%%%%%%%%%%%%%%%%%%%%%%

\vspace{10mm}

\end{document}